# The ESO Key-Programme "A Homogeneous Bright QSO Survey" - I [*]

## The Methods and the "Deep" Fields

S. Cristiani[1], F. La Franca[1], P. Andreani[1], A. Gemmo[2], P. Goldschmidt[3], L. Miller[4], R. Vio[1,5], C. Barbieri[1], L. Bodini[1], A. Iovino[6], M. Lazzarin[1], R. Clowes[7], H. MacGillivray[4], Ch. Gouiffes[8], C. Lissandrini[1], and A. Savage[4,9]

[1] Dipartimento di Astronomia, Vicolo dell'Osservatorio 5, I-35122 Padova, Italy
[2] European Southern Observatory, K. Schwarzschild Strasse 2, D-85748 Garching, Germany
[3] Physics Department, Imperial College, London SW7 2BZ, UK
[4] Royal Observatory, Blackford Hill, Edinburgh EH9 3HJ, Scotland
[5] ESA, IUE Observatory, Villafranca del Castillo, Apartado 50727, 28080 Madrid, Spain
[6] Osservatorio Astronomico di Brera, via Brera 28, I-20121 Milano, Italy
[7] Centre for Astrophysics, Univ. of Central Lancashire, Preston PR1 2HE, UK
[8] CEA,DSM/DAPNIA/Service d'Astrophysique C.E Saclay 91191 Gif sur Yvette Cedex, France
[9] UK Schmidt Telescope, Anglo-Australian Observatory, Coonabarabran, New South Wales 2357, Australia

**Abstract.** This is the first paper in a series aimed at defining a statistically significant sample of QSOs in the range $15 < B < 18.75$ and $0.3 < z < 2.2$. The selection is carried out using direct plates obtained at the ESO and UK Schmidt Telescopes, scanned with the COSMOS facility and searched for objects with an ultraviolet excess. Follow-up spectroscopy, carried out at ESO La Silla, is used to classify each candidate. In this initial paper, we describe the scientific objectives of the survey; the selection and observing techniques used. We present the first sample of 285 QSOs ($M_B < -23$) in a 153 deg$^2$ area, covered by the six "deep" fields, intended to obtain significant statistics down $B \simeq 18.75$ with unprecedented photometric accuracy. From this database, QSO counts are determined in the magnitude range $17 < B < 18.75$.

## 1. Introduction

QSO surveys provide basic information to address a number of cosmological key issues: the epoch of galaxy forma-tion, the physics driving the AGN phenomenon, the formation and evolution of the large scale structure, the UV and X-ray backgrounds. In this respect, with the development of techniques for automatic detection and analysis of images, a powerful tool has become available to astronomers. However, to obtain useful cosmological information, it is necessary to carry out statistically well-defined surveys, quantifying the selection effects in the first place and optimizing the investment of telescope time for spectroscopic confirmation of the candidates.

Any uncertainty or systematic bias of the databases affects the determination of the shape of the Luminosity Function (LF) and its evolution. Understanding the type of luminosity evolution followed by individual objects may indicate the mechanism that fuels the central engine, revealing if the QSO phenomenon is driven by the surrounding environment or determined by its nuclear conditions only. Ascertaining how well a pure luminosity evolution describes the data may help in discriminating a "long-lived" against a "recurrent activity" model for QSOs. Results in one sense or another can be spuriously favoured by fits that overlook the observational biases. To probe the real trend not only the database at faint magnitudes has to be enlarged but also the incompleteness at bright magnitudes should be bound or removed with better samples allowing an adequately detailed analysis.

The PG bright QSO survey (Schmidt & Green 1983) and the MBQS (Mitchell et al. 1984), at the bright end

**Table 1.** List of the observed fields

| Field | Center | | Limits | | | | Area | Selection Interval | Selection Type |
|---|---|---|---|---|---|---|---|---|---|
| | $\alpha$ (1950) | $\delta$ (1950) | $\Delta\alpha_{min}$ (deg) | $\Delta\alpha_{max}$ (deg) | $\Delta\delta_{min}$ (deg) | $\Delta\delta_{max}$ (deg) | (deg$^2$) | | |
| 287 | 21 28 03.0 | -44 54 51 | 21 15 29 | 21 40 37 | -2.202 | +2.202 | 19.93 | $16.00 < B_J \le 18.50$ | $U - B_J \;\; B_J - R$ |
| 295 | 00 50 15.7 | -40 01 59 | -2.640 | +2.600 | -2.603 | +2.603 | 27.09 | $15.00 < B_J \le 18.50$ | $U - B_J \;\; B_J - V'$ |
| 351 | 00 46 38.8 | -35 07 23 | -2.721 | +2.560 | -2.678 | +2.678 | 27.98 | $15.50 < B_J \le 18.50$ | $U - B_J \;\; B_J - V'$ |
| 534 | 22 44 09.6 | -24 58 48 | -2.538 | +2.560 | -2.500 | +2.500 | 25.07 | $15.50 < B' \le 18.25$ | $U - B' \;\; B' - V'$ |
| SA94 | 02 48 23.7 | -00 08 51 | -2.613 | +2.713 | -2.632 | +2.632 | 27.95 | $14.00 < B_J \le 18.85$ | $U - B_J \;\; B_J - R$ |
| SGP | 00 50 34.3 | -28 10 08 | -2.170 | +2.715 | -2.520 | +2.520 | 24.56 | $15.00 < B_J \le 18.70$ | $U - B_J \;\; B_J - R$ |
| | | | | | | | 152.58 | | |

of the QSO counts, are especially affected because of the shape of the luminosity function and the relatively large photometric errors, compared with other existing samples at fainter magnitudes. A true increase of information about LF shape and evolution is obtained only combining rich "homogeneous" samples, i.e. with sensibly matched signal-to-noise ratios: if anything, because of the Bennett effect (Murdoch, Crawford & Jauncey, 1973), the signal-to-noise ratio ought to be larger for the brighter surveys, which sample the steeper part of the LF.

## 2. The Homogeneous Bright QSO Survey (HBQS)

From the above considerations we have been prompted to undertake a new survey with high photometric accuracy, required to investigate the bright part of the counts, where the PG sample is incomplete and the information from MBQS is scanty.

In order to obtain a statistically significant number of QSOs, an area of the sky of several hundred deg$^2$ has to be surveyed. We have selected the zone around the south galactic pole ($b < -60°$), which has the following advantages:

1. the galactic absorption is minimized. It may be argued, in fact, that the imperfect knowledge of the galactic extinction prevents a determination of a photometric scale better than a tenth of a magnitude and, ultimately, a detailed understanding of the fluctuations of the QSO surface density (Weedman, 1986). Furthermore, at lower galactic latitudes the reddening acts to loosen the colour-selection criteria. Even a small modification would bring to the QSO locus a number of galactic candidates, which would be very expensive in terms of telescope time;

2. the contamination from galactic objects is minimized;

3. a number of other surveys, with complementary techniques, have been carried out on this area of the sky, providing an easy way to calibrate our selection criteria and check the effectiveness and completeness of our search.

In 1989 ESO has endorsed as a Key-Programme this project to create an high-quality QSO sample in the magnitude range $15 < B < 18.5$. In this paper we describe in detail the procedure followed in the data acquisition and reduction and report the first results obtained in six fields ($\sim 150$ deg$^2$, the ones reaching fainter limiting magnitudes), providing significant statistics in the range ($16.75 < B < 18.75$). Forthcoming papers, enlarging the surveyed area, will address the issue of QSO counts at brighter magnitudes ($15 < B < 17$).

Table 1 gives the list of the observed fields. For each of them the coordinates of the center are reported together with the limits of the area over which the selection has been carried out. Except for field 287 (see the individual notes below) the area investigated is approximately a rectangle on the sky, whose dimensions in degrees are also given. Unusable areas (due to the presence of bright stars or galaxies, plate defects, etc.) are reported in the individual notes.

## 3. Data Acquisition and Reduction

Two Schmidt plates for each bandpass $U$, $B'$ (or $B_J$), $V'$, $R$ (or $OR$), $I$ (see Blair & Gilmore 1982 for a definition of the various systems) have been obtained at the UKST or ESO, usually within a few months interval, in order to minimize the effects of variability. The plates used are listed field by field in Table 2.

The plate material has been scanned on the COS-MOS microdensitometer (Mac Gillivray & Stobie 1984). The resulting tables, containing the instrumental magnitudes and other useful parameters for the objects detected in each plate, have been merged together in one table per field. Only objects with at least 4 detections (in 10 plates) have been included in this final table. The astrometric error box defining a common detection has been defined as a circle of 3.5 arcsec of radius. In this way plate defects are easily rejected and spurious detections are minimized to an acceptable level. In addition to the usual $\alpha$ and $\delta$, a further angular coordinate, $RAC$, has been defined as

**Table 2.** Plates used for the Selection of the Candidates

### Field 287 UK-Schmidt

| emulsion | filter | band and number | date | exp. time (min.) |
|---|---|---|---|---|
| IIIa-J | UG1 | U10294 | 1985-06-25 | 240 |
| IIIa-J | GG395 | J10260 | 1985-06-14 | 70 |
| IIIa-J | GG395 | J10264 | 1985-06-15 | 70 |
| IIIa-J | GG395 | J10271 | 1985-06-16 | 70 |
| IIIa-J | GG395 | J10292 | 1985-06-24 | 90 |
| IIIa-F | RG630 | R10302 | 1985-06-26 | 100 |
| IIIa-F | RG630 | R10306 | 1985-06-27 | 105 |
| IIIa-F | RG630 | R10341 | 1985-07-20 | 105 |
| IIIa-F | RG630 | R10349 | 1985-07-23 | 90 |

### Field 295 UK-Schmidt

| emulsion | filter | band and number | date | exp. time (min.) |
|---|---|---|---|---|
| IIIa-J | UG1 | U13249 | 1989-08-25 | 180 |
| IIIa-J | UG1 | U13253 | 1989-08-26 | 180 |
| IIIa-J | GG395 | J12310 | 1987-12-14 | 95 |
| IIIa-J | GG395 | J12312 | 1987-12-16 | 95 |
| IIIa-J | GG395 | SRCJ295 | 1986-11-22 | 75 |
| IIa-D | GG495 | V13223 | 1989-08-03 | 180 |
| IIa-D | GG495 | V13239 | 1989-08-29 | 60 |
| IIIa-F | OG590 | OR11404 | 1986-09-27 | 50 |
| IIIa-F | OG590 | OR13222 | 1988-08-03 | 70 |
| IIIa-F | OG590 | OR13870 | 1990-09-24 | 75 |
| IV-N | RG715 | I12282 | 1987-11-26 | 90 |
| IV-N | RG715 | I12285 | 1987-11-27 | 90 |

### Field 351 UK-Schmidt

| emulsion | filter | band and number | date | exp. time (min.) |
|---|---|---|---|---|
| IIIa-J | UG1 | U12114 | 1987-09-02 | 180 |
| IIIa-J | UG1 | U12158 | 1987-09-17 | 225 |
| IIIa-J | GG395 | J12095 | 1987-08-25 | 140 |
| IIIa-J | GG395 | J12216 | 1987-10-14 | 39 |
| IIa-D | GG495 | V12152 | 1987-09-16 | 60 |
| IIa-D | GG495 | V12186 | 1987-09-20 | 60 |
| IIIa-F | OG590 | OR12105 | 1987-08-26 | 35 |
| IIIa-F | OG590 | OR12226 | 1987-10-20 | 100 |
| IV-N | RG715 | I12009 | 1987-07-17 | 90 |
| IV-N | RG715 | I12107 | 1987-01-09 | 90 |

### Field 534 ESO-La-Silla

| emulsion | filter | band and number | date | exp. time (min.) |
|---|---|---|---|---|
| IIa-O | UG1 | U8196 | 1989-08-31 | 75 |
| IIa-O | UG1 | U8197 | 1989-08-31 | 75 |
| IIa-O | GG385 | B8194 | 1989-08-31 | 30 |
| IIa-O | GG385 | B8198 | 1989-08-31 | 30 |
| 103a-D | GG495 | V8260 | 1989-10-20 | 30 |
| 103a-D | GG495 | V8261 | 1989-10-20 | 30 |
| 103a-D | GG495 | V8710 | 1990-08-23 | 30 |
| IIIa-F | RG630 | R8257 | 1989-10-19 | 40 |
| IIIa-F | RG630 | R8258 | 1989-10-19 | 40 |
| IV-N | RG715 | I8221 | 1989-09-16 | 60 |
| IV-N | RG715 | I8222 | 1989-09-18 | 60 |

### Field SA94 UK-Schmidt

| emulsion | filter | band and number | date | exp. time (min.) |
|---|---|---|---|---|
| IIIa-J | UG1 | U12780 | 1988-10-09 | 240 |
| IIIa-J | UG1 | U13396 | 1989-10-21 | 180 |
| IIIa-J | GG395 | J12227 | 1987-10-20 | 95 |
| IIIa-J | GG395 | J13408 | 1989-10-28 | 70 |
| IIa-D | GG495 | V12813 | 1988-10-17 | 60 |
| IIa-D | GG495 | V13415 | 1989-10-30 | 60 |
| IIIa-F | RG630 | R12218 | 1987-10-16 | 150 |
| IIIa-F | RG630 | R12807 | 1988-10-15 | 150 |
| IV-N | RG715 | I12812 | 1988-10-17 | 90 |
| IV-N | RG715 | I13392 | 1989-10-20 | 90 |

### Field SGP UK-Schmidt

| emulsion | filter | band and number | date | exp. time (min.) |
|---|---|---|---|---|
| IIa-O | UG1 | U2639 | 1976-12-12 | 180 |
| IIIa-J | UG1 | U6326 | 1980-09-03 | 180 |
| IIIa-J | UG1 | U6326 | 1980-09-15 | 180 |
| IIIa-J | GG395 | J9764 | 1984-11-23 | 95 |
| IIIa-J | GG395 | J9765 | 1984-11-23 | 70 |
| IIIa-J | GG395 | J9766 | 1984-11-23 | 70 |
| IIIa-J | GG395 | J9770 | 1984-11-24 | 70 |
| IIIa-J | GG395 | J9771 | 1984-11-24 | 70 |
| IIIa-F | RG630 | R4676 | 1978-12-02 | 90 |
| IIIa-F | RG630 | R9594 | 1984-09-22 | 90 |
| IIIa-F | RG630 | R9672 | 1984-10-17 | 90 |
| IIIa-F | RG630 | R11336 | 1986-09-02 | 90 |

**Table 3.** Photometric Errors

| Band | Coefficient Order | | |
|------|:---:|:---:|:---:|
| | 0 | 1 | 2 |
| **Field 287** | | | |
| $U$ | 0.07 | ... | ... |
| $B_J$ | 0.07 | ... | ... |
| $R$ | 0.07 | ... | ... |

| Band | Coefficient Order | | |
|------|:---:|:---:|:---:|
| | 0 | 1 | 2 |
| **Field 295** | | | |
| $U$ | 0.07 | ... | ... |
| $B_J$ | 0.07 | ... | ... |
| $V'$ | 5.00 | $-0.609$ | $1.881 \cdot 10^{-2}$ |
| $R$ | 0.07 | ... | ... |

| Band | Coefficient Order | | |
|------|:---:|:---:|:---:|
| | 0 | 1 | 2 |
| **Field 351** | | | |
| $U$ | $-0.02$ | 0.007 | ... |
| $B_J$ | $-1.38$ | 0.158 | $-4.15 \cdot 10^{-3}$ |
| $V'$ | 0.07 | ... | ... |
| $OR$ | 0.07 | ... | ... |

| Band | Coefficient Order | | |
|------|:---:|:---:|:---:|
| | 0 | 1 | 2 |
| **Field 534** | | | |
| $U$ | 0.07 | ... | ... |
| $B'$ | 0.07 | ... | ... |
| $V'$ | $-0.49$ | 0.0392 | ... |
| $R$ | 0.08 | ... | ... |

| Band | Coefficient Order | | |
|------|:---:|:---:|:---:|
| | 0 | 1 | 2 |
| **Field SA 94** | | | |
| $U$ | 0.07 | ... | ... |
| $B_J$ | 0.07 | ... | ... |
| $V'$ | 0.78 | $-0.1045$ | $3.8175 \cdot 10^{-3}$ |
| $R$ | 0.08 | ... | ... |

| Band | Coefficient Order | | |
|------|:---:|:---:|:---:|
| | 0 | 1 | 2 |
| **Field SGP** | | | |
| $U$ | 0.07 | ... | ... |
| $B_J$ | 0.07 | ... | ... |
| $R$ | 0.07 | ... | ... |

$$RAC = (\alpha - \alpha_{center}) \cdot 15 \cdot \cos(\delta)$$

$RAC$ represents the right ascension distance (in decimal degrees) on the sky with respect to the field center. The limits of the fields reported in Table 1 have then been defined in terms of $RAC$ and $\delta$.

The calibrated magnitudes and the photometric errors have been derived in the following way:

1. for each plate the histogram of the instrumental magnitudes has been analysed to find the limit of com-
the histogram;

2. for each bandpass, the quantity $\Delta mag$ has been computed as the difference of the instrumental magnitudes of the plate whose errors are to be determined minus the instrumental magnitudes of the other plate (or the averages of the other plates);

3. for each plate, the objects, ordered in ascending instrumental magnitude, have been subdivided into groups of 1001, and the mean magnitude for each group has been determined;

4. for each group, the objects have been ordered in ascending $\Delta mag$. If the mean magnitude is less than the limit-magnitude, the distribution of the $\Delta mag$'s is expected to be gaussian, centered around a certain value, corresponding to the $501^{th}$ object. The difference of the $\Delta mag$'s corresponding to the $842^{th}$ and the $158^{th}$ objects gives twice the $\sigma$ of the distribution. For magnitudes fainter than the limit-magnitude, the mode of the distribution of the $\Delta mag$'s has been determined and only the part of the histogram with $\Delta mag$'s less than the mode has been analysed to obtain an indication of the variance.

5. the errors for the given plate as a function of the mean instrumental magnitude of the group have then been derived from the above estimates of the $\sigma$'s with the assumption that all the plates are of the same quality;

6. then the regression of the errors has been evaluated for each plate as a function of the instrumental magnitude;

7. the relationship between calibrated and instrumental magnitudes has been derived for each bandpass via a least square fitting that takes into account the errors both on the instrumental and calibrating magnitudes and treats simultaneously all the plates of that bandpass;

8. the resulting photometric uncertainties of the calibrated magnitudes as a function of the $B_J$ or $B'$ magnitude have been fitted with low-order polynomials whose coefficients are reported in Table 3;

9. the colors of main sequence stars were used as a standard reference: shifts in the position of the main stellar locus on a two-color diagram, measured against the corresponding distribution at the center of the plate, were used to estimate and correct position-dependent and magnitude-dependent systematic photometric errors.

The present careful assessment of the photometric properties of the survey not only enables us to obtain a reliable and efficient selection of the QSO candidates (see below), but will also allow a detailed estimate of the selection function in forthcoming papers.

To separate extended from point-like sources, we have computed the difference between the measured FWHM of a given object and the mode of the FWHM distribution at the magnitude of the object. This quantity has then been averaged on the six pairs of plates B (or J), V, R (or

**Table 4.** List of the Candidates

### Field 287

| $\alpha$ (1950) | $\delta$ (1950) | $B_J$ | $U - B_J$ | $B_J - R$ | Id. | z | Ref. |
|---|---|---|---|---|---|---|---|
| 21 15 58.80 | −45 42 31.0 | 17.50 | −0.42 | 0.94 | s | 0.000 | |
| 21 16 29.94 | −44 00 08.9 | 17.13 | −0.33 | 0.06 | s | 0.000 | |
| 21 16 55.06 | −44 39 37.7 | 18.22 | −0.31 | 0.72 | Q | 1.480 | 2 |
| 21 17 07.54 | −44 16 38.1 | 16.09 | −0.52 | −0.08 | s | 0.000 | |
| 21 17 30.02 | −44 02 41.1 | 18.35 | −0.59 | 0.50 | Q | 1.710 | 2 |
| 21 19 26.28 | −42 49 39.8 | 18.50 | −0.59 | 0.73 | Q | 1.054 | 2 |
| 21 20 50.84 | −43 27 56.3 | 18.20 | −0.57 | 0.61 | Q | 1.240 | 2 |
| 21 21 35.76 | −42 57 51.9 | 17.92 | −0.53 | −0.05 | s | 0.000 | |
| 21 21 41.94 | −44 06 58.2 | 17.92 | −0.62 | 0.53 | Q | 1.735 | 2 |
| 21 21 51.46 | −42 57 00.7 | 17.51 | −0.15 | 0.16 | s | 0.000 | |
| 21 22 00.50 | −45 57 25.1 | 17.68 | −0.67 | 0.76 | Q | 0.9534 | |
| 21 23 21.24 | −43 38 30.0 | 18.43 | −0.40 | 0.37 | Q | 0.481 | |
| 21 25 32.64 | −46 43 50.5 | 16.25 | −0.40 | 0.70 | s | 0.000 | |
| 21 26 10.06 | −42 56 31.3 | 18.14 | −0.47 | 0.53 | Q | 0.4103 | |
| 21 26 40.94 | −43 34 19.8 | 18.44 | −0.55 | 0.09 | Q | 0.584 | 2 |
| 21 26 48.48 | −46 52 34.4 | 18.26 | −0.68 | 0.42 | s | 0.000 | |
| 21 27 19.38 | −42 42 44.8 | 17.72 | −0.84 | 0.33 | Q | 0.799 | 2 |
| 21 27 56.48 | −45 13 49.9 | 18.49 | −0.56 | 0.97 | s | 0.000 | |
| 21 28 13.18 | −43 27 06.6 | 17.49 | −0.56 | 0.33 | Q | 0.920 | 2 |
| 21 28 21.94 | −46 10 53.5 | 17.49 | −0.60 | 0.26 | Q | 0.835 | |
| 21 29 09.68 | −43 24 52.9 | 18.22 | −0.94 | 0.34 | Q | 1.984 | |
| 21 29 39.48 | −46 24 19.6 | 18.12 | −0.39 | 0.48 | Q | 0.435 | 2 |
| 21 30 24.84 | −43 57 43.5 | 18.20 | −0.39 | 0.37 | s | 0.000 | |
| 21 31 33.46 | −42 57 51.3 | 17.61 | −0.78 | 0.05 | Q | 2.093 | 2 |
| 21 31 59.68 | −44 49 51.9 | 17.78 | −0.43 | 0.52 | ... | ... | |
| 21 32 13.52 | −45 16 50.4 | 17.88 | −0.52 | 0.46 | Q | 0.5067 | |
| 21 33 51.08 | −44 08 27.9 | 17.98 | −0.09 | 0.05 | s | 0.000 | |
| 21 34 50.90 | −46 01 49.9 | 18.13 | −0.27 | 0.93 | Q | 0.528 | 2 |
| 21 35 19.46 | −46 20 48.0 | 18.39 | −0.15 | 0.33 | Q | 0.505 | 2 |
| 21 36 30.54 | −43 01 22.9 | 18.04 | −0.90 | 0.80 | Q | 1.343 | 2 |
| 21 36 46.70 | −43 44 38.3 | 18.24 | −0.41 | 0.77 | Q | 0.4852 | |
| 21 37 35.42 | −45 00 13.9 | 17.13 | −0.51 | 0.32 | s | 0.000 | |
| 21 37 51.76 | −44 35 49.2 | 18.49 | −0.55 | 0.26 | Q | 0.630 | |
| 21 38 41.52 | −46 50 34.1 | 18.25 | −0.26 | 0.16 | Q | 0.762 | 2 |
| 21 40 10.20 | −45 42 30.1 | 16.30 | −0.60 | 0.38 | Q | 0.171 | 1 |
| 21 40 16.54 | −45 52 37.7 | 18.35 | −0.46 | 0.32 | Q | 1.688 | 2 |

OR) and empirically used to distinguish between point-like and extended objects. At the magnitudes of interest for the present work, such a separation is not a critical point, due to the relatively low percentage of extended objects, therefore only extremely extended objects (for which in any case the quality of the photometry is doubtful) have been excluded from the subsequent selection of the QSO candidates. We have selected as candidates all the UVx "not-extremely-extended" objects satisfying a type of *modified Braccesi less-restricted* two-color criterion (cfr. Braccesi et al. 1980, La Franca et al. 1992). The adopted two-color selections are listed for each field in Table 1. The principal color for the selection has been chosen to be the $U - B'$ or the $U - B_J$ according to the type of the available plate material. The secondary color has been chosen to be $B_J - R$ or $B_J - V'$ or $B' - V'$, always preferring the combination giving the most reliable photometric accuracy.

In the case of a two-color $U - B'$, $B' - V'$ diagram the criterion can be expressed by the following formula:

$$\text{for } 0.45 \leq B' - V' \leq 0.80 : \quad U - B' < -0.40;$$
$$\text{for } B' - V' < 0.45 :$$
$$\sqrt{[(U - B') - 0.1]^2 + [(B' - V') - 0.45]^2} > 0.5 \quad \text{and}$$
$$U - B' < -0.15$$

**Table 4.** - continued

## Field 295

| $\alpha$ (1950) | $\delta$ (1950) | $B_J$ | $U - B_J$ | $B_J - V'$ | $B_J - OR$ | Id. | $z$ | Ov | Ref |
|---|---|---|---|---|---|---|---|---|---|
| 0 36 38.43 | -39 54 00.2 | 17.30 | -1.70 | -0.02 | -0.13 | s | 0.000 | | |
| 0 36 59.84 | -42 01 19.1 | 16.48 | -0.82 | 0.19 | 0.28 | s | 0.000 | | |
| 0 37 17.92 | -39 45 42.3 | 18.47 | -0.19 | 0.21 | 0.53 | Q | 1.464 | | |
| 0 37 34.99 | -37 58 38.4 | 18.05 | -0.04 | -1.07 | -0.08 | Q | 1.772 | | |
| 0 37 42.17 | -40 41 24.3 | 16.45 | -1.47 | -0.15 | -0.21 | s | 0.000 | | |
| 0 37 46.11 | -39 32 54.7 | 18.30 | -0.08 | -0.31 | 0.11 | Q | 0.473 | | |
| 0 37 46.87 | -41 48 56.8 | 17.32 | -0.79 | 0.15 | 0.55 | Q | 1.273 | | |
| 0 37 49.67 | -42 15 46.3 | 15.96 | -0.59 | 0.45 | 0.77 | s | 0.000 | | |
| 0 38 15.16 | -38 38 07.3 | 16.79 | -0.77 | 0.22 | 0.67 | Q | 0.135 | | |
| 0 38 32.73 | -39 02 45.4 | 18.25 | -0.70 | 0.01 | 0.36 | Q | 0.704 | | |
| 0 38 33.20 | -37 38 56.9 | 17.14 | -0.69 | 0.32 | 0.34 | s | 0.000 | 351 | |
| 0 38 34.73 | -37 48 44.3 | 18.05 | -0.26 | 0.37 | 0.79 | s | 0.000 | | |
| 0 38 39.22 | -37 26 07.0 | 17.26 | -0.79 | 0.35 | 0.48 | s | 0.000 | 351 | |
| 0 39 20.32 | -39 30 52.6 | 17.78 | -0.80 | -0.06 | 0.39 | Q | 1.624 | | |
| 0 39 22.37 | -39 47 59.7 | 18.16 | -1.07 | 0.19 | 0.48 | Q | 1.466 | | |
| 0 39 35.66 | -39 38 13.1 | 18.12 | -0.98 | 0.50 | 0.87 | Q | 1.848 | | |
| 0 39 43.40 | -40 22 01.0 | 18.38 | -0.79 | 0.32 | 0.56 | Q | 1.933 | | |
| 0 40 01.26 | -37 29 48.6 | 18.37 | -0.84 | -0.08 | -0.16 | s | 0.000 | 351 | |
| 0 40 32.90 | -37 31 56.9 | 18.18 | -0.69 | 0.39 | 0.94 | Q | 1.780 | 351 | |
| 0 40 34.31 | -37 58 09.7 | 15.77 | -0.25 | 0.30 | 0.35 | s | 0.000 | | |
| 0 40 37.89 | -39 29 55.5 | 18.04 | -0.38 | 0.00 | 0.34 | s | 0.000 | | |
| 0 41 13.60 | -37 50 18.6 | 18.03 | -0.43 | 0.34 | 0.72 | Q | 1.070 | | |
| 0 41 55.47 | -38 04 27.9 | 18.28 | -0.20 | 0.05 | 0.02 | s | 0.000 | | |
| 0 42 03.83 | -38 43 59.6 | 18.16 | -0.45 | 0.21 | 0.29 | Q | 1.562 | | |
| 0 42 04.56 | -42 32 38.8 | 15.72 | -0.95 | 0.12 | 0.08 | s | 0.000 | | |
| 0 42 12.03 | -41 36 09.6 | 18.27 | -0.60 | 0.05 | -0.01 | s | 0.000 | | |
| 0 42 32.30 | -41 35 36.9 | 18.28 | -0.91 | -0.10 | -0.13 | s | 0.000 | | |
| 0 42 58.57 | -42 03 43.7 | 17.74 | -0.12 | -0.41 | 0.15 | Q | 0.469 | | |
| 0 43 12.25 | -37 41 07.2 | 17.86 | -0.47 | 0.13 | 0.05 | s | 0.000 | 351 | |
| 0 43 17.21 | -39 24 07.2 | 15.70 | -0.80 | 0.31 | 0.58 | s | 0.000 | | |
| 0 43 24.29 | -41 43 38.3 | 18.30 | -1.39 | -0.38 | -0.14 | s | 0.000 | | |
| 0 43 37.66 | -38 40 32.1 | 15.56 | -1.57 | -0.06 | -0.16 | s | 0.000 | | |
| 0 43 48.99 | -42 11 27.7 | 18.25 | -0.15 | -0.29 | 0.02 | Q | 0.926 | | |
| 0 43 51.33 | -40 52 26.1 | 15.38 | -1.04 | 0.17 | 0.12 | s | 0.000 | | |
| 0 44 13.35 | -40 59 11.1 | 17.03 | -1.37 | 0.00 | -0.08 | s | 0.000 | | |
| 0 44 18.53 | -41 16 36.5 | 18.40 | -0.21 | 0.03 | 0.34 | s | 0.000 | | |
| 0 44 22.57 | -37 26 45.6 | 17.99 | -0.31 | ... | 0.51 | Q | 0.308 | 351 | |
| 0 44 54.64 | -41 51 37.1 | 18.23 | -0.37 | 0.04 | 0.34 | s | 0.000 | | |
| 0 45 34.80 | -41 10 16.2 | 18.18 | -0.66 | 0.14 | 0.77 | Q | 1.354 | | |
| 0 45 48.02 | -40 21 55.1 | 18.34 | -0.54 | 0.23 | 0.71 | Q | 1.637 | | |
| 0 45 48.54 | -38 26 54.6 | 18.41 | -1.40 | 0.76 | 1.22 | Q | 0.400 | | 2 |
| 0 46 03.75 | -41 43 34.5 | 18.42 | -0.89 | 0.21 | 0.65 | Q | 1.880 | | |
| 0 46 08.14 | -38 28 24.7 | 18.48 | -0.31 | -0.09 | 0.86 | Q | 0.442 | | |
| 0 46 08.23 | -39 46 22.5 | 18.41 | -0.29 | 0.21 | 0.29 | s | 0.000 | | |
| 0 46 38.76 | -38 30 10.0 | 15.07 | -0.37 | 0.52 | 0.84 | s | 0.000 | | |
| 0 46 40.44 | -39 06 17.6 | 18.28 | -0.79 | 0.21 | 0.60 | Q | 1.964 | | |
| 0 46 44.68 | -40 35 13.3 | 17.79 | -0.41 | -0.41 | 0.28 | s | 0.000 | | |
| 0 47 10.55 | -37 51 57.0 | 17.60 | -0.97 | -0.18 | -0.17 | s | 0.000 | | |
| 0 47 11.45 | -40 04 43.7 | 18.27 | -0.91 | 0.06 | -0.06 | s | 0.000 | | |
| 0 48 19.58 | -39 40 36.9 | 17.22 | -1.16 | 0.11 | 0.29 | Q | 0.482 | | |
| 0 48 42.57 | -38 33 42.9 | 17.07 | -1.03 | -0.34 | 0.02 | s | 0.000 | | |
| 0 49 10.55 | -38 47 33.6 | 17.75 | -0.70 | -0.10 | 0.45 | Q | 0.395 | | |
| 0 49 31.52 | -41 16 11.8 | 18.13 | -0.35 | 0.06 | 0.12 | s | 0.000 | | |
| 0 49 51.77 | -39 48 55.0 | 15.70 | -0.51 | 0.35 | 0.62 | s | 0.000 | | |

**Table 4.** - continued

## Field 295 - end

| $\alpha$ (1950) | $\delta$ (1950) | $B_J$ | $U - B_J$ | $B_J - V'$ | $B_J - OR$ | Id. | $z$ | Ov | Ref |
|---|---|---|---|---|---|---|---|---|---|
| 0 49 52.90 | -40 42 47.3 | 17.36 | -0.42 | 0.66 | 1.06 | s | 0.000 | | |
| 0 51 24.47 | -41 57 38.2 | 17.28 | -0.17 | 0.09 | 0.12 | s | 0.000 | | |
| 0 51 48.37 | -39 33 08.2 | 16.93 | -0.01 | -0.37 | 0.31 | Q | 0.224 | | |
| 0 52 03.26 | -37 30 53.1 | 17.47 | -1.18 | 0.14 | 0.47 | nq | 0.000 | 351 | |
| 0 52 08.40 | -40 50 41.0 | 18.43 | -0.87 | -0.09 | 0.09 | s | 0.000 | | |
| 0 52 09.49 | -41 04 01.1 | 18.17 | -0.93 | 0.21 | 0.55 | Q | 2.077 | | |
| 0 52 21.94 | -39 28 11.3 | 18.35 | -0.35 | 0.25 | 0.40 | s | 0.000 | | |
| 0 52 25.90 | -38 26 21.9 | 18.43 | -0.19 | 0.24 | 0.67 | Q | 2.541 | | |
| 0 52 47.02 | -40 49 07.3 | 17.47 | -0.91 | 0.20 | 0.18 | s | 0.000 | | |
| 0 52 47.95 | -42 33 36.2 | 17.82 | -0.35 | 0.20 | 0.63 | s | 0.000 | | |
| 0 52 50.07 | -40 20 05.3 | 18.26 | -0.72 | 0.26 | 0.73 | Q | 1.239 | | |
| 0 53 00.19 | -41 46 27.1 | 17.75 | -0.53 | 0.33 | 1.05 | s | 0.000 | | |
| 0 53 12.90 | -38 26 34.5 | 18.09 | -0.63 | -0.16 | 0.53 | Q | 0.379 | | 2 |
| 0 53 26.82 | -37 58 02.0 | 18.35 | -0.45 | -1.23 | -0.60 | ext | ... | | |
| 0 54 37.43 | -39 02 32.2 | 15.44 | -0.43 | 0.27 | 0.60 | s | 0.000 | | |
| 0 54 46.43 | -38 54 13.4 | 16.05 | -0.49 | -0.05 | 0.06 | s | 0.000 | | |
| 0 55 10.21 | -42 14 48.0 | 18.32 | -0.82 | 0.28 | 1.00 | Q | 0.276 | | |
| 0 55 17.84 | -42 24 43.9 | 15.57 | -0.47 | 0.39 | 0.68 | s | 0.000 | | |
| 0 55 24.04 | -39 25 06.9 | 17.79 | -0.40 | 0.65 | 1.02 | s | 0.000 | | |
| 0 55 28.67 | -39 55 36.6 | 16.05 | -0.40 | -0.03 | 0.08 | s | 0.000 | | |
| 0 55 29.87 | -39 27 16.9 | 17.19 | -0.81 | 0.06 | -0.02 | s | 0.000 | | |
| 0 55 43.08 | -37 48 11.2 | 18.12 | -0.09 | -0.46 | -0.11 | s | 0.000 | | |
| 0 56 21.02 | -39 24 52.1 | 17.56 | -0.92 | 0.36 | 0.68 | Q | 1.415 | | |
| 0 56 34.62 | -38 01 04.6 | 18.43 | -0.59 | 0.15 | 0.75 | Q: | 0.875 | | |
| 0 57 06.15 | -41 26 55.3 | 17.27 | -0.72 | 0.14 | 0.28 | s | 2.065 | | |
| 0 57 09.21 | -42 24 03.5 | 18.13 | -0.27 | 0.16 | 0.51 | s | 0.000 | | |
| 0 57 56.66 | -42 19 48.9 | 18.10 | -0.71 | 0.47 | 0.70 | Q | 0.672 | | |
| 0 58 12.34 | -40 29 59.8 | 17.90 | -0.47 | -0.18 | 0.14 | Q | 0.610 | | |
| 0 58 22.05 | -41 34 40.9 | 17.17 | -0.55 | 0.13 | 0.65 | s | 0.000 | | |
| 0 58 39.53 | -40 08 54.7 | 17.81 | -1.28 | 0.10 | 0.08 | s | 0.000 | | |
| 0 59 40.83 | -41 09 59.1 | 17.90 | -1.08 | 0.51 | 0.76 | Q | 1.963 | | |
| 1 00 12.67 | -41 03 02.5 | 18.37 | -0.30 | 0.27 | 0.60 | Q: | 1.032 | | |
| 1 00 19.38 | -42 20 10.7 | 16.67 | -0.87 | -0.03 | -0.03 | s | 0.000 | | |
| 1 00 30.84 | -40 23 31.8 | 18.34 | -0.70 | 0.03 | -0.06 | s | 0.000 | | |
| 1 00 40.64 | -40 49 39.8 | 18.43 | -0.23 | 0.03 | 0.40 | s | 0.000 | | |
| 1 00 47.54 | -42 20 12.0 | 17.03 | -1.04 | 0.13 | 0.42 | Q | 1.895 | | |
| 1 00 57.41 | -40 53 14.6 | 15.65 | -0.24 | 0.17 | 0.40 | s | 0.000 | | |
| 1 01 28.56 | -40 59 05.3 | 18.46 | -0.92 | 0.35 | 0.80 | Q | 0.363 | | |
| 1 02 03.78 | -37 43 36.7 | 17.08 | -1.09 | 0.03 | 0.11 | s | 0.000 | | |
| 1 02 30.58 | -39 48 49.2 | 18.36 | -0.02 | -0.07 | 0.18 | Q | 1.540 | | |
| 1 02 36.98 | -37 55 26.7 | 17.67 | -0.58 | -0.01 | 0.00 | s | 0.000 | | |
| 1 02 37.43 | -37 32 13.7 | 18.43 | -0.66 | 0.27 | 0.58 | Q | 0.380 | | 2 |
| 1 02 38.40 | -39 00 17.7 | 18.47 | -0.12 | 0.00 | 0.16 | Q | 0.594 | | |
| 1 03 20.19 | -41 34 00.1 | 18.12 | -0.29 | 0.20 | 0.48 | Q | 0.778 | | |
| 1 03 23.58 | -41 02 07.4 | 17.62 | -0.77 | -1.89 | -1.45 | s | 0.000 | | |

## Field 351

| α (1950) | δ (1950) | $B_J$ | $U - B_J$ | $B_J - V$ | $B_J - OR$ | Id. | z | Ov. | ref. |
|---|---|---|---|---|---|---|---|---|---|
| 00 33 45.04 | -35 24 00.1 | 18.27 | -1.13 | 0.16 | 0.34 | Q | 0.409 | | |
| 00 33 46.82 | -34 40 05.9 | 16.48 | -0.64 | 0.07 | 0.20 | s | 0.000 | | |
| 00 33 51.24 | -35 31 43.5 | 18.27 | -1.05 | -0.05 | -0.15 | nq | 0.000 | | |
| 00 33 56.21 | -32 27 22.8 | 17.71 | -0.70 | 0.52 | 0.71 | Q | 2.174 | | |
| 00 34 11.55 | -33 09 04.7 | 17.91 | -0.78 | 0.28 | 0.47 | Q | 2.180 | | 2 |
| 00 34 16.50 | -33 20 58.2 | 17.60 | -0.35 | 0.56 | 0.84 | s | 0.000 | | |
| 00 34 48.41 | -35 26 46.6 | 17.55 | -0.36 | 0.47 | 0.70 | s | 0.000 | | |
| 00 35 35.06 | -34 17 51.2 | 16.70 | -0.44 | 0.55 | 0.81 | s | 0.000 | | |
| 00 35 39.23 | -35 18 10.7 | 18.06 | -0.89 | 0.49 | 0.85 | Q | 1.199 | | |
| 00 35 48.04 | -32 50 52.4 | 18.27 | -0.57 | 0.11 | 0.53 | Q | 0.458 | | |
| 00 35 51.14 | -35 20 34.9 | 18.38 | -0.54 | 0.04 | 0.37 | Q | 1.519 | | |
| 00 35 53.77 | -33 23 46.2 | 18.26 | -0.94 | 0.36 | 0.54 | Q | 1.042 | | |
| 00 35 53.86 | -34 38 29.4 | 18.44 | -0.76 | 0.17 | 0.58 | Q | 1.335 | | |
| 00 36 02.12 | -36 40 55.9 | 17.51 | -0.62 | 0.05 | 0.37 | Q | 0.403 | | |
| 00 36 02.98 | -35 25 49.4 | 18.13 | -0.42 | 0.64 | 1.04 | nq | 0.000 | | |
| 00 36 20.57 | -34 26 56.6 | 18.33 | -0.61 | 0.13 | 0.34 | Q | 0.840 | | |
| 00 36 33.30 | -35 36 04.6 | 18.36 | -0.96 | 0.09 | 0.28 | Q | 1.853 | | |
| 00 36 44.96 | -35 30 00.8 | 17.99 | -0.59 | 0.23 | 0.41 | s | 0.000 | | |
| 00 37 04.09 | -37 08 02.7 | 18.41 | -0.72 | 0.39 | 0.47 | Q | 0.811 | | |
| 00 37 11.35 | -35 45 45.5 | 18.49 | -0.56 | 0.01 | 0.78 | Q | 1.095 | | |
| 00 37 16.67 | -34 36 09.7 | 18.45 | -0.60 | 0.16 | 0.47 | Q | 1.599 | | |
| 00 37 16.68 | -35 44 29.6 | 18.09 | -0.56 | 0.23 | 0.54 | Q | 0.836 | | |
| 00 37 38.39 | -33 27 10.1 | 18.08 | -0.71 | 0.28 | 0.75 | Q | 1.172 | | |
| 00 37 39.37 | -34 23 39.9 | 18.41 | -0.78 | -0.02 | -0.07 | s | 0.000 | | |
| 00 37 59.08 | -33 36 11.9 | 17.88 | -0.73 | -0.18 | -0.21 | s | 0.000 | | |
| 00 38 18.02 | -34 56 24.7 | 17.29 | -0.67 | 0.38 | 0.74 | HII | 0.199 | | |
| 00 38 33.22 | -37 38 59.9 | 17.08 | -0.31 | 0.32 | 0.34 | s | 0.000 | 295 | |
| 00 38 39.24 | -37 26 10.1 | 17.18 | -0.47 | 0.31 | 0.44 | s | 0.000 | 295 | |
| 00 38 58.30 | -35 16 18.5 | 18.39 | -0.40 | 0.21 | 0.39 | s | 0.000 | | |
| 00 39 02.42 | -36 13 20.9 | 18.27 | -0.66 | 0.20 | 0.53 | Q | 1.468 | | |
| 00 39 05.68 | -34 12 46.5 | 18.07 | -0.51 | 0.10 | 0.21 | Q | 0.585 | | |
| 00 39 50.51 | -33 54 21.5 | 18.34 | -0.47 | 0.30 | 0.67 | Q | 2.480 | | |
| 00 40 01.26 | -37 29 51.5 | 18.32 | -0.76 | -0.04 | -0.23 | s | 0.000 | 295 | |
| 00 40 03.02 | -33 42 31.4 | 18.47 | -0.47 | 0.46 | 0.91 | s | 0.000 | | |
| 00 40 19.06 | -37 04 08.1 | 18.03 | -0.16 | 0.18 | 0.36 | Q | 2.720 | | |
| 00 40 32.93 | -37 31 59.6 | 18.11 | -0.65 | 0.46 | 0.81 | Q | 1.780 | 295 | |
| 00 41 13.20 | -34 34 11.6 | 18.48 | -0.76 | 0.27 | 0.65 | Q | 1.478 | | |
| 00 42 36.30 | -36 24 41.5 | 17.03 | -0.73 | 0.15 | 0.81 | Q | 0.410 | | |
| 00 42 53.12 | -32 49 53.0 | 18.50 | -0.50 | 0.23 | 0.67 | Q | 1.283 | | |
| 00 43 03.64 | -36 44 44.6 | 18.46 | -0.75 | 0.19 | 0.77 | Q | 1.930 | | |
| 00 43 12.28 | -37 41 09.9 | 17.98 | -0.42 | 0.00 | 0.01 | s | 0.000 | 295 | |
| 00 43 17.87 | -36 55 56.2 | 17.31 | -0.64 | 0.09 | 1.04 | Q | 0.326 | | |
| 00 43 42.47 | -36 36 56.2 | 17.93 | -0.72 | -0.15 | -0.15 | s | 0.000 | | |
| 00 44 22.59 | -37 26 48.4 | 18.23 | -0.81 | 0.40 | 0.76 | nq | 0.308 | | |
| 00 44 28.80 | -35 54 00.7 | 17.87 | -0.45 | 0.11 | 0.06 | s | 0.000 | | |
| 00 44 34.14 | -32 51 40.3 | 17.61 | -0.47 | 0.13 | 0.51 | Q | 1.573 | | |
| 00 44 59.88 | -34 14 37.8 | 17.83 | -0.62 | 0.31 | 0.51 | Q | 0.875 | | |
| 00 46 05.74 | -35 30 00.7 | 18.30 | -0.64 | 0.34 | 0.38 | Q | 2.170 | | 2 |
| 00 46 11.22 | -33 54 04.2 | 17.94 | -0.72 | 0.27 | 0.59 | Q | 1.155 | | |
| 00 46 17.05 | -36 21 39.6 | 17.39 | -0.55 | 0.00 | 0.10 | s | 0.000 | | |
| 00 47 06.30 | -34 10 25.3 | 16.63 | -0.13 | -0.05 | 0.03 | s | 0.000 | | |
| 00 47 33.16 | -32 52 55.4 | 17.82 | -0.86 | -0.10 | -0.10 | s | 0.000 | | |
| 00 47 35.78 | -36 41 16.6 | 18.44 | -0.63 | -0.05 | 0.19 | s | 0.000 | | |
| 00 49 19.51 | -37 18 03.2 | 18.37 | -0.38 | 0.21 | 0.70 | HII | 0.029 | | |

## Field 351 - end

| α (1950) | δ (1950) | $B_J$ | $U - B_J$ | $B_J - V$ | $B_J - OR$ | Id. | z | Ov. | ref. |
|---|---|---|---|---|---|---|---|---|---|
| 00 49 26.02 | -33 21 09.3 | 18.26 | -0.36 | 0.38 | 0.84 | HII | 0.217 | | |
| 00 49 38.82 | -34 12 04.8 | 18.46 | -0.77 | 0.17 | 0.63 | Q | 1.369 | | |
| 00 49 43.30 | -33 18 03.4 | 18.36 | -0.58 | 0.33 | 0.61 | Q | 0.824 | | |
| 00 49 47.77 | -36 31 14.9 | 17.29 | -0.53 | 0.21 | 0.36 | s | 0.000 | | |
| 00 49 54.15 | -35 34 14.0 | 18.25 | -0.47 | 0.47 | 0.79 | Q | 1.478 | | |
| 00 50 17.85 | -34 57 38.5 | 18.13 | -0.82 | 0.21 | 0.55 | Q | 1.440 | | |
| 00 50 19.83 | -34 13 12.4 | 18.08 | -1.10 | -0.22 | -0.28 | s | 0.000 | | |
| 00 50 51.76 | -35 09 51.4 | 18.20 | -0.67 | 0.40 | 0.79 | Q | 1.480 | | |
| 00 51 21.59 | -33 26 20.0 | 18.29 | -0.41 | 0.39 | 0.83 | Q | 1.633 | | |
| 00 51 39.87 | -36 33 34.2 | 18.36 | -0.64 | 0.11 | 0.30 | nq | ... | | |
| 00 51 49.21 | -36 48 39.6 | 18.41 | -0.61 | 0.15 | 0.03 | s | 0.000 | | |
| 00 52 02.69 | -34 01 51.2 | 16.97 | -0.94 | -0.07 | -0.08 | s | 0.000 | | |
| 00 52 03.28 | -37 30 55.2 | 17.41 | -0.84 | 0.12 | 0.37 | nq | 0.000 | 295 | |
| 00 52 48.23 | -36 37 31.4 | 15.84 | -0.72 | 0.19 | 0.35 | s | 0.000 | | |
| 00 53 11.26 | -35 00 50.8 | 17.92 | -0.22 | 0.27 | 0.38 | Q | 2.300 | | |
| 00 54 02.14 | -34 03 56.2 | 17.59 | -0.95 | -0.14 | -0.17 | s | 0.000 | | |
| 00 55 49.44 | -33 54 02.7 | 18.22 | -0.48 | 0.37 | 0.46 | Q | 0.832 | | |
| 00 56 15.75 | -36 22 16.3 | 15.50 | -0.97 | 0.30 | 0.63 | Q | 0.162 | | 2 |
| 00 56 15.79 | -34 39 44.2 | 18.11 | -0.67 | 0.16 | 0.35 | Q | 0.525 | | |
| 00 56 19.64 | -34 25 19.8 | 18.44 | -0.51 | 0.36 | 0.63 | Q | 1.525 | | |
| 00 56 34.27 | -35 57 14.4 | 18.33 | -0.42 | 0.32 | 0.76 | Q | 2.340 | | |
| 00 56 49.61 | -33 25 41.2 | 17.46 | -0.90 | 0.18 | 0.50 | Q | 1.815 | | |
| 00 56 52.54 | -35 57 52.6 | 18.40 | -0.40 | 0.36 | 0.57 | Q | 2.730 | | |
| 00 57 02.67 | -34 20 56.9 | 18.36 | -0.50 | 0.20 | 0.51 | Q | 0.564 | | |
| 00 57 29.31 | -32 31 54.1 | 17.89 | -0.69 | 0.02 | 0.41 | Q | 0.533 | | |
| 00 57 51.85 | -35 50 06.4 | 17.97 | -0.49 | 0.10 | 0.48 | Q | 1.587 | | |
| 00 58 08.57 | -34 22 42.8 | 18.18 | -0.44 | 0.27 | 0.74 | Q | 1.727 | | |
| 00 58 16.30 | -32 31 05.0 | 17.26 | -0.60 | 0.08 | 0.57 | Q | 1.574 | | |
| 00 58 22.80 | -34 13 20.2 | 18.48 | -0.63 | 0.41 | 0.47 | Q | 0.878 | | |

In the case of other two-color selections the criterion has been adjusted to correspond to the above formula for typical colors of main sequence stars. For example in the case of a $U - B_J$, $B_J - R$ diagram the following criterion has been adopted:

$$\text{for } 0.75 \leq B_J - R \leq 1: \quad U - B_J < -0.40;$$

$$\text{for } B_J - R < 0.75:$$

$$\sqrt{[(U - B_J) - 0.55]^2 + [(B_J - R) - 0.75]^2} > 0.9 \quad \text{and}$$

$$U - B_J < -0.15$$

In all the fields the reliability of the adopted two-color selection has been checked against other criteria: for example the $U - B_J$ vs $B_J - R$ selections have been compared with the $U - B_J$ vs $B_J - V'$ and with the $U - B_J$ vs $2B_J - V' - R$ diagrams; multi-color criteria (based on a power-law fit of all the available broad-band photometric data) have also been used. In this way the $U - B'$ or the $U - B_J$ selection limits have been occasionally shifted (always within $\pm 0.05$ mag) to take into account possible systematic zero point errors in the $U$ calibration. Addi-

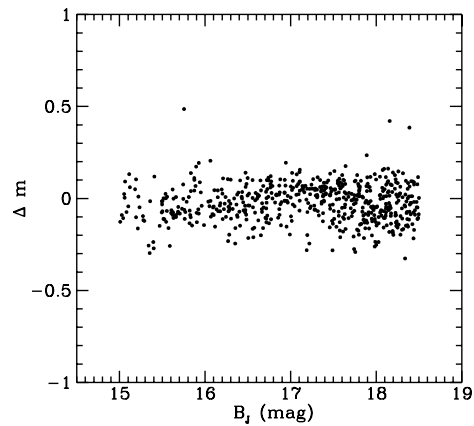

**Fig. 1.** The magnitude difference $\Delta m = B_J(295) - B_J(351)$ for point-like objects as a function of the $B_J(295)$ magnitude in the interval $15 < B_J < 18.5$

**Table 4.** - continued

## Field 534

| $\alpha$ (1950) | $\delta$ (1950) | $B'$ | $U - B'$ | $B' - V'$ | $B' - R$ | Id. | $z$ |
|---|---|---|---|---|---|---|---|
| 22 33 36.33 | -24 03 19.7 | 17.84 | -0.42 | -0.12 | 0.18 | Q | 1.505 |
| 22 33 58.42 | -26 48 15.8 | 18.06 | -0.43 | 0.02 | 0.65 | Q | 1.641 |
| 22 34 00.00 | -24 36 30.3 | 17.63 | -0.79 | 0.08 | 0.27 | Q | 1.849 |
| 22 35 13.01 | -26 46 47.3 | 16.69 | -1.19 | -0.44 | -0.64 | s | 0.000 |
| 22 35 31.53 | -23 01 44.2 | 17.48 | -0.75 | 0.33 | 0.51 | Q | 0.568 |
| 22 35 38.86 | -23 27 08.0 | 17.17 | -1.07 | -0.26 | -0.38 | nq | 0.529 |
| 22 36 23.22 | -24 52 23.2 | 17.71 | -0.42 | 0.23 | 0.65 | Q | 0.529 |
| 22 36 45.48 | -24 52 18.4 | 17.92 | -0.35 | 0.18 | 0.15 | s | 0.000 |
| 22 37 30.64 | -25 05 02.2 | 16.90 | -1.18 | -0.23 | -0.27 | s | 0.000 |
| 22 38 18.11 | -24 45 01.5 | 17.93 | -0.74 | 0.24 | 0.22 | Q | 0.854 |
| 22 38 46.42 | -24 27 25.7 | 17.33 | -1.18 | -0.46 | -0.62 | s | 0.000 |
| 22 39 54.93 | -26 41 45.1 | 16.22 | -1.56 | 0.54 | ... | s: | 0.000 |
| 22 40 03.74 | -27 21 12.7 | 18.12 | -0.57 | 0.13 | 0.00 | Q | 0.764 |
| 22 40 21.34 | -22 58 17.9 | 15.97 | -1.09 | -0.37 | -0.65 | s: | 0.000 |
| 22 40 40.03 | -26 38 45.0 | 15.71 | -1.16 | -0.11 | -0.32 | s | 0.000 |
| 22 40 57.25 | -24 11 01.5 | 17.12 | -0.58 | 0.05 | 0.47 | Q | 0.184 |
| 22 41 06.77 | -25 16 12.0 | 17.53 | -0.52 | 0.11 | 0.38 | Q | 0.455 |
| 22 41 56.78 | -24 18 48.5 | 16.95 | -0.89 | 0.12 | 0.30 | Q | 1.958 |
| 22 42 01.41 | -25 21 47.5 | 17.86 | -0.65 | 0.03 | 0.20 | Q | 0.319 |
| 22 43 18.73 | -27 01 15.5 | 18.10 | -0.46 | 0.10 | 0.43 | s | 0.000 |
| 22 43 20.62 | -25 20 59.8 | 17.72 | -0.77 | 0.37 | 0.56 | Q | 1.274 |
| 22 43 35.33 | -24 38 41.2 | 18.05 | -0.85 | 0.28 | 0.59 | Q | 1.405 |
| 22 43 36.74 | -23 01 48.3 | 18.19 | -0.49 | 0.22 | 0.29 | s | 0.000 |
| 22 43 48.91 | -23 46 56.7 | 16.81 | -0.52 | 0.52 | 0.80 | s | 0.000 |
| 22 44 08.70 | -24 34 53.6 | 17.71 | -0.58 | 0.10 | 0.08 | s | 0.000 |
| 22 44 57.87 | -24 39 46.1 | 18.23 | -0.97 | 0.05 | 0.35 | Q | 0.736 |
| 22 45 13.23 | -25 35 40.5 | 16.53 | -0.19 | -0.10 | 0.08 | s | 0.000 |
| 22 45 44.20 | -23 46 35.6 | 17.40 | -0.65 | 0.24 | 0.44 | Q | 1.026 |
| 22 45 47.63 | -26 13 32.2 | 18.22 | -0.55 | 0.09 | -0.21 | s | 0.000 |
| 22 46 40.35 | -24 55 47.1 | 17.62 | -0.84 | 0.73 | 1.09 | g | 0.050 |
| 22 46 52.96 | -23 42 03.1 | 17.76 | -0.23 | -0.12 | -0.14 | s | 0.000 |
| 22 46 55.61 | -27 22 47.1 | 15.87 | -1.29 | -0.03 | -0.03 | s | 0.000 |
| 22 47 26.83 | -26 44 05.2 | 15.72 | -0.70 | -0.14 | -0.38 | s | 0.000 |
| 22 47 42.72 | -25 04 21.0 | 17.74 | -0.92 | -0.06 | 0.21 | Q | 2.075 |
| 22 48 13.61 | -27 07 32.8 | 18.03 | -0.65 | 0.23 | 0.49 | Q | 0.820 |
| 22 49 02.78 | -24 23 40.4 | 17.77 | -0.76 | -0.24 | 0.36 | Q | 1.688 |
| 22 49 13.60 | -22 29 59.9 | 17.20 | -0.80 | 0.38 | 0.61 | Q | 0.219 |
| 22 50 08.03 | -24 43 35.1 | 18.21 | -0.19 | -0.26 | 0.00 | Q | 1.718 |
| 22 50 21.34 | -27 04 06.5 | 17.88 | -0.73 | -0.27 | -0.48 | s | 0.000 |
| 22 51 21.02 | -26 18 12.2 | 18.01 | -0.73 | -0.08 | 0.44 | Q | 1.016 |
| 22 51 35.55 | -25 21 00.7 | 17.83 | -0.83 | 0.18 | 0.34 | Q | 1.341 |
| 22 52 42.63 | -24 34 30.7 | 16.33 | -0.80 | 0.40 | 0.74 | Q | 0.147 |
| 22 53 09.35 | -22 42 48.0 | 18.25 | -0.46 | -0.19 | -0.14 | Q | 0.563 |
| 22 53 23.70 | -23 29 49.9 | 17.62 | -0.67 | -0.04 | -0.11 | s | 0.000 |
| 22 53 52.42 | -23 55 26.8 | 18.09 | -0.52 | -0.18 | 0.16 | Q | 1.374 |
| 22 54 02.93 | -24 55 34.4 | 17.23 | -0.59 | 0.32 | 0.18 | s | 0.000 |
| 22 54 12.79 | -25 09 59.9 | 17.50 | -0.21 | -0.22 | 0.83 | s | 0.000 |
| 22 54 23.58 | -24 47 24.6 | 17.06 | -0.77 | 0.34 | 0.41 | Q | 1.091 |
| 22 54 39.72 | -25 39 00.8 | 17.05 | -0.44 | 0.65 | 0.91 | s | 0.000 |
| 22 55 12.12 | -26 26 41.2 | 17.90 | -0.60 | 0.38 | 0.43 | s | 0.000 |

**Table 4.** - continued

## Field SA94

| α (1950) | δ (1950) | $B_J$ | $U - B_J$ | $B_J - V'$ | $B_J - R$ | Id. | z | Ref. |
|---|---|---|---|---|---|---|---|---|
| 2 38 03.96 | -0 01 30.1 | 18.84 | -0.58 | 0.07 | 0.78 | Q | 0.468 | |
| 2 38 13.96 | 2 16 31.0 | 17.50 | -1.16 | 0.36 | 0.50 | Q | 1.900 | |
| 2 38 19.78 | 0 54 07.7 | 18.72 | -0.35 | 0.51 | 0.78 | ... | ... | |
| 2 38 25.55 | -0 52 23.2 | 17.91 | -0.81 | -0.21 | 0.07 | s | 0.000 | |
| 2 38 33.85 | 1 54 51.2 | 18.31 | -0.57 | 0.44 | 0.84 | Q | 1.973 | |
| 2 38 40.13 | -1 42 04.9 | 17.45 | -0.83 | 0.02 | 0.65 | Q | 0.346 | |
| 2 38 47.00 | 1 17 18.7 | 18.61 | -0.39 | -0.03 | 0.16 | g | ... | |
| 2 38 51.33 | -2 45 02.7 | 17.64 | -0.78 | 0.26 | 0.70 | s | 0.000 | |
| 2 38 53.53 | -0 58 13.7 | 18.46 | -0.56 | -0.08 | 0.02 | Q | 0.726 | |
| 2 39 07.70 | -0 08 29.6 | 18.52 | -0.59 | 0.04 | 0.03 | Q | 0.649 | |
| 2 39 21.95 | 0 21 56.7 | 18.36 | -0.57 | 0.31 | 0.49 | Q | 1.054 | |
| 2 39 23.29 | -0 05 41.8 | 18.72 | -0.62 | 0.25 | 0.62 | Q | 1.552 | |
| 2 39 27.14 | -0 12 24.2 | 18.70 | -0.60 | 0.00 | 0.37 | Q | 1.112 | |
| 2 39 47.49 | 0 36 28.7 | 18.48 | -0.31 | -0.03 | 0.50 | s | 0.000 | |
| 2 40 05.85 | 0 44 43.9 | 16.60 | -0.73 | 0.19 | 0.41 | Q | 0.569 | 2 |
| 2 40 17.88 | 1 31 26.5 | 18.51 | -1.06 | -0.14 | -0.32 | s | 0.000 | |
| 2 40 30.91 | -0 12 37.0 | 18.71 | -0.86 | -0.08 | 0.39 | Q | 2.018 | |
| 2 40 36.21 | -2 10 31.2 | 18.04 | -1.14 | 0.05 | 0.61 | Q | 1.899 | 2 |
| 2 40 36.96 | -2 20 48.7 | 17.67 | -0.79 | 0.37 | 0.43 | s | 0.000 | |
| 2 40 40.55 | 1 45 35.5 | 18.68 | -0.66 | -0.50 | 0.51 | Q | 2.063 | |
| 2 40 41.00 | 1 07 15.2 | 16.80 | -0.96 | 0.04 | 0.01 | s | 0.000 | |
| 2 41 39.85 | 0 05 54.5 | 18.84 | -0.65 | 0.01 | 0.24 | Q | 0.684 | |
| 2 41 49.74 | -1 55 21.1 | 17.11 | -1.09 | 0.00 | 0.24 | s | 0.000 | |
| 2 41 53.72 | -1 33 45.8 | 18.81 | -0.27 | -0.29 | 0.47 | Q | 0.2015 | |
| 2 41 55.21 | 2 15 35.1 | 18.36 | -0.18 | -0.13 | 0.45 | HII | 0.073 | |
| 2 41 55.77 | -2 14 09.7 | 17.91 | -1.65 | 0.69 | 1.07 | s: | 0.000 | |
| 2 42 13.73 | -2 11 15.6 | 18.84 | -0.46 | 0.06 | 0.26 | Q | 0.680 | |
| 2 42 15.74 | -0 40 15.6 | 18.81 | -0.45 | -0.02 | 0.41 | s | 0.000 | |
| 2 42 24.00 | -2 20 59.3 | 18.18 | -0.67 | 0.37 | 0.96 | HII | 0.161 | |
| 2 42 24.83 | 1 12 44.0 | 18.63 | -0.42 | ... | 0.69 | ... | ... | |
| 2 42 59.28 | 2 11 29.2 | 18.42 | -0.39 | 0.40 | 0.88 | s | 0.000 | |
| 2 43 17.79 | -0 59 20.8 | 17.43 | -0.21 | 0.16 | 0.32 | s | 0.000 | |
| 2 43 28.29 | -1 12 11.7 | 18.67 | -0.57 | 0.38 | 0.96 | Q | 0.201 | |
| 2 43 28.54 | 0 13 05.0 | 17.44 | -0.87 | 0.05 | 0.01 | s | 0.000 | |
| 2 43 30.35 | -0 44 45.9 | 18.69 | -0.67 | 0.18 | 0.44 | Q | 1.596 | 2 |
| 2 43 36.81 | 1 10 50.2 | 18.58 | -0.64 | 0.06 | 0.61 | Q | 1.591 | |
| 2 43 43.41 | -0 18 36.0 | 18.52 | -0.74 | 0.15 | 0.52 | Q | 1.139 | |
| 2 43 56.85 | -2 04 24.2 | 18.49 | -0.75 | -0.20 | -0.07 | s | 0.000 | |
| 2 43 58.86 | -2 39 51.5 | 15.76 | -0.52 | 0.38 | 0.61 | s | 0.000 | |
| 2 44 01.99 | -0 21 23.2 | 18.64 | -0.80 | -0.04 | 0.32 | Q | 1.817 | 2 |
| 2 44 06.50 | 0 58 59.8 | 18.15 | -0.53 | 0.61 | 1.36 | s | 0.000 | |
| 2 44 14.82 | -1 58 07.2 | 18.26 | -0.60 | 0.12 | 0.46 | Q | 1.784 | 2 |
| 2 44 18.96 | -1 12 03.0 | 17.16 | -0.78 | 0.32 | 0.59 | Q | 0.467 | 2 |
| 2 44 25.50 | -0 58 42.6 | 18.26 | -0.62 | ... | 0.71 | ... | ... | |
| 2 45 12.43 | -0 08 57.2 | 16.45 | -1.03 | 0.04 | -0.18 | s | 0.000 | |
| 2 45 22.81 | -0 28 24.7 | 18.55 | -0.83 | 0.27 | 0.45 | Q | 2.118 | 2 |
| 2 45 47.58 | 0 38 42.0 | 18.23 | -0.71 | 0.11 | -0.01 | s | 0.000 | |
| 2 45 47.81 | -2 34 28.2 | 18.00 | -0.32 | 0.86 | -0.09 | ... | ... | |
| 2 46 05.11 | 2 11 16.8 | 18.70 | -0.75 | 0.34 | 0.78 | Q | 1.267 | |
| 2 46 54.24 | 0 57 00.5 | 18.73 | -0.82 | 0.03 | 0.47 | Q | 0.954 | 2 |
| 2 47 05.46 | -2 34 20.6 | 18.40 | -1.02 | 0.20 | 0.54 | Q | 1.935 | 2 |
| 2 47 06.93 | -1 24 15.1 | 18.76 | -0.05 | -0.23 | -0.21 | HII | 0.068 | |

## Field SA94 - continued

| $\alpha$ (1950) | $\delta$ (1950) | $B_J$ | $U - B_J$ | $B_J - V'$ | $B_J - R$ | Id. | $z$ | Ref. |
|---|---|---|---|---|---|---|---|---|
| 2 47 32.97 | 0 13 03.2 | 17.88 | -0.83 | 0.29 | 0.52 | Q | 0.1984 | |
| 2 47 57.16 | -0 20 23.2 | 18.16 | -0.57 | 0.11 | 0.56 | Q | 1.458 | 2 |
| 2 47 59.84 | -1 10 44.4 | 17.55 | -0.88 | 0.21 | 0.00 | s | 0.000 | |
| 2 48 05.58 | -1 00 00.1 | 18.70 | -1.05 | 0.13 | 0.56 | Q | 1.845 | 2 |
| 2 48 14.91 | -0 10 12.8 | 18.81 | -0.42 | 0.17 | 0.53 | Q | 0.766 | 2 |
| 2 48 30.00 | 1 07 09.0 | 18.37 | -0.60 | 0.31 | 1.07 | Q | 0.232 | |
| 2 48 32.53 | -2 21 09.3 | 18.68 | -0.84 | 0.10 | 0.37 | Q | 1.516 | |
| 2 48 50.59 | -1 25 33.4 | 18.63 | -1.15 | -0.37 | 0.15 | s | 0.000 | |
| 2 49 11.13 | -2 34 24.3 | 17.49 | -0.44 | 0.41 | 0.87 | ... | ... | |
| 2 49 16.82 | 1 44 33.7 | 16.10 | -0.26 | -0.21 | 0.85 | HII | 0.0752 | |
| 2 49 21.84 | 0 44 49.6 | 18.85 | -0.58 | 0.07 | 0.89 | Q | 0.470 | 2 |
| 2 49 26.63 | 0 33 27.4 | 18.51 | -0.71 | 0.04 | 0.14 | s | 0.000 | |
| 2 49 29.66 | -1 17 32.5 | 16.72 | -1.17 | 0.34 | 0.76 | s | 0.000 | |
| 2 49 42.40 | 2 22 56.8 | 18.65 | -0.60 | 0.08 | 0.37 | Q | 2.805 | |
| 2 49 42.55 | -2 24 42.7 | 18.84 | -0.23 | -0.20 | 0.21 | HII | 0.0385 | |
| 2 49 47.22 | -0 06 16.6 | 17.45 | -0.58 | 0.31 | 0.52 | Q | 0.810 | 2 |
| 2 49 53.84 | -1 58 00.8 | 18.47 | -0.49 | 0.40 | 0.79 | s | 0.000 | |
| 2 50 13.30 | -1 44 05.0 | 18.48 | -0.46 | 0.12 | 0.97 | s | 0.000 | |
| 2 50 15.38 | -2 17 15.5 | 18.60 | -0.35 | 0.16 | 0.44 | Q | 1.116 | |
| 2 50 19.30 | -2 37 33.1 | 14.63 | -0.61 | 0.24 | -0.02 | s | 0.000 | |
| 2 50 29.14 | -1 42 20.2 | 17.46 | -1.06 | -0.14 | -0.01 | s | 0.000 | |
| 2 50 29.60 | 1 04 24.2 | 18.46 | 0.06 | 0.24 | 0.69 | s | 0.000 | |
| 2 50 30.83 | 0 01 33.8 | 17.55 | -0.61 | 0.33 | 0.42 | s | 0.000 | |
| 2 50 52.40 | -0 40 04.3 | 18.25 | -0.84 | -0.07 | -0.07 | s | 0.000 | |
| 2 51 06.48 | -0 51 41.2 | 18.65 | -0.62 | -0.01 | -0.07 | s | 0.000 | |
| 2 51 07.07 | -0 01 01.8 | 18.76 | -0.58 | 0.13 | 0.77 | Q | 1.677 | |
| 2 51 29.25 | 0 46 43.5 | 17.29 | -1.34 | -0.23 | -0.40 | s | 0.000 | |
| 2 51 52.97 | -0 53 32.3 | 16.86 | -0.38 | -0.30 | 0.87 | HII | 0.0148 | |
| 2 51 53.50 | -2 41 11.8 | 18.72 | -0.92 | -0.19 | -0.22 | s | 0.000 | |
| 2 51 56.86 | 1 11 01.8 | 18.03 | -0.34 | -0.35 | 0.65 | HII | 0.028 | |
| 2 51 59.37 | -0 54 29.2 | 18.06 | -0.15 | -0.22 | 0.54 | Q | 0.433 | 2 |
| 2 52 07.14 | -1 53 22.5 | 16.35 | -0.21 | -0.20 | 0.64 | HII | 0.0381 | |
| 2 52 08.10 | 1 41 09.3 | 18.00 | -0.44 | 0.32 | 0.65 | Q | 0.620 | |
| 2 52 31.63 | 0 13 15.5 | 18.12 | -0.74 | -0.09 | 0.88 | Q | 0.354 | 2 |
| 2 52 31.90 | -2 17 51.7 | 15.85 | -0.51 | 0.45 | 0.82 | s | 0.000 | |
| 2 52 40.05 | 1 36 21.2 | 18.04 | -0.41 | 0.18 | 0.64 | Q | 2.457 | |
| 2 52 51.92 | 2 25 08.1 | 17.35 | -0.68 | 0.14 | 0.25 | s | 0.000 | |
| 2 52 58.64 | 1 18 56.1 | 17.14 | -0.54 | 0.20 | 0.83 | Q | 0.141 | |
| 2 53 03.52 | 1 02 34.1 | 18.70 | -0.28 | 0.01 | 0.44 | s | 0.000 | |
| 2 53 25.49 | 0 41 06.6 | 18.69 | -0.98 | 0.06 | 0.23 | Q | 0.847 | 2 |
| 2 53 43.72 | -0 09 06.5 | 18.39 | -0.34 | 0.43 | 0.67 | ... | ... | |
| 2 53 44.01 | -1 38 42.1 | 16.92 | -0.89 | 0.22 | 0.37 | Q | 0.878 | |
| 2 53 45.90 | -0 57 05.1 | 18.82 | -0.70 | 0.24 | 0.46 | Q | 0.720 | |
| 2 53 50.54 | 0 23 54.6 | 18.19 | -1.32 | -0.19 | -0.28 | s | 0.000 | |
| 2 53 51.15 | 0 22 25.7 | 18.58 | -0.45 | -0.24 | 0.73 | HII | 0.013 | |
| 2 54 10.81 | 0 00 43.6 | 18.26 | -0.55 | 0.19 | 0.57 | Q | 2.242 | |
| 2 54 27.07 | -2 35 24.7 | 18.27 | -0.77 | 0.29 | 0.47 | Q | 1.106 | |
| 2 54 34.01 | -2 01 27.8 | 15.97 | -0.57 | 0.42 | 0.75 | s | 0.000 | |
| 2 54 34.68 | 0 34 26.8 | 17.58 | -0.76 | 0.07 | 0.02 | s | 0.000 | |
| 2 54 41.97 | -2 45 07.9 | 17.95 | -0.53 | 0.45 | 0.62 | HII | 0.176 | |
| 2 54 44.28 | -1 46 19.7 | 14.62 | -0.43 | 0.47 | 0.53 | s | 0.000 | |
| 2 54 59.25 | -1 54 50.4 | 18.71 | -0.75 | 0.09 | 0.06 | s | 0.000 | |

**Field SA94 - end**

| α (1950) | δ (1950) | $B_J$ | $U - B_J$ | $B_J - V'$ | $B_J - R$ | Id. | z | Ref. |
|---|---|---|---|---|---|---|---|---|
| 2 55 11.79 | -1 58 39.8 | 15.32 | -0.58 | 0.45 | 0.71 | s | 0.000 | |
| 2 55 11.87 | 0 49 06.5 | 17.74 | -0.91 | -0.12 | -0.20 | s | 0.000 | |
| 2 55 13.62 | -1 31 46.5 | 18.57 | -0.82 | 0.30 | 0.72 | Q | 1.520 | |
| 2 55 20.69 | 1 32 37.1 | 18.61 | -0.91 | 0.34 | 0.94 | Q | 0.282 | |
| 2 55 28.22 | -1 05 59.0 | 18.18 | -0.65 | 0.09 | 0.29 | s | 0.000 | |
| 2 55 43.21 | 0 57 47.8 | 18.38 | -1.14 | 0.16 | 0.60 | s | 0.000 | |
| 2 55 45.88 | 1 59 44.5 | 16.78 | -1.38 | -0.29 | -0.25 | s | 0.000 | |
| 2 55 50.09 | -1 58 00.9 | 14.42 | -0.56 | 0.56 | 0.92 | s | 0.000 | |
| 2 56 03.89 | -2 01 18.1 | 16.07 | -0.44 | 0.37 | 0.66 | s | 0.000 | |
| 2 56 05.94 | -2 06 06.8 | 18.59 | -1.01 | 0.30 | 0.54 | Q | 0.406 | |
| 2 56 12.41 | 1 32 47.3 | 18.20 | 0.10 | -0.28 | 0.04 | s | 0.000 | |
| 2 56 14.68 | 1 40 28.3 | 18.77 | -0.57 | 0.38 | 0.41 | Q | 0.608 | 2 |
| 2 56 36.99 | -0 34 34.6 | 18.11 | -0.88 | 0.04 | 0.75 | Q | 0.361 | |
| 2 56 38.91 | -1 29 48.2 | 18.65 | -0.54 | -0.03 | 0.14 | s | 0.000 | |
| 2 56 55.08 | -0 31 53.9 | 17.74 | -1.28 | 0.16 | 0.69 | Q | 1.998 | |
| 2 56 56.45 | -1 34 22.1 | 18.63 | -0.97 | 0.15 | -0.07 | s | 0.000 | |
| 2 57 00.46 | -0 37 10.9 | 18.63 | -0.54 | 0.32 | 1.15 | Q | 1.748 | 2 |
| 2 57 02.60 | -1 25 58.6 | 17.44 | -1.59 | -0.30 | -0.24 | s | 0.000 | |
| 2 57 03.24 | 0 25 42.6 | 16.87 | -0.84 | 0.33 | 0.64 | Q | 0.535 | |
| 2 57 43.11 | 1 16 45.9 | 18.81 | -0.84 | 0.10 | 0.80 | Q | 1.356 | 2 |
| 2 57 51.19 | -0 35 32.9 | 17.00 | -1.55 | -0.14 | -0.16 | s | 0.000 | |
| 2 58 02.47 | -0 27 23.3 | 18.48 | -0.79 | 0.18 | 0.50 | Q | 1.435 | 2 |
| 2 58 08.16 | 2 27 50.6 | 18.64 | -0.70 | 0.36 | 0.65 | Q: | 0.892 | |
| 2 58 10.37 | 2 10 53.5 | 17.99 | -0.55 | 0.17 | 0.57 | Q | 2.521 | |
| 2 58 14.48 | 0 42 50.5 | 18.72 | -0.66 | 0.28 | -0.02 | Q | 0.661 | 2 |
| 2 58 20.48 | -1 29 43.2 | 18.37 | -0.57 | 0.30 | 0.66 | HII | 0.0027 | |
| 2 58 25.65 | 1 37 38.5 | 18.80 | -0.37 | -0.10 | 0.42 | Q | 0.595 | 2 |
| 2 58 36.11 | -2 39 41.0 | 18.53 | -0.27 | 0.54 | 0.69 | ... | ... | |
| 2 58 56.76 | 0 49 34.3 | 18.13 | -0.73 | -0.06 | 0.60 | s | 0.000 | |
| 2 59 04.00 | -2 32 50.3 | 18.71 | -0.37 | 0.96 | ... | ... | ... | |

tional candidates outside the selection criteria have also been observed to further test the presence of systematic errors, but no new QSO with redshift below 2.2 has been found.

An easy way to test the reliability of the procedure of photometric calibration and error estimation is to check the magnitudes of the objects in the area of sky in common between the two fields 295 and 351 (a comparison in an unfavorable situation, since this region is at the edges of the plates, where the field effects are expected to be at their maximum).

In Fig. 1 the magnitude difference $B_J(295) - B_J(351)$ is plotted for the point-like objects as a function of the $B_J(295)$ magnitude in the interval of selection $15 < B_J < 18.5$. The results agrees perfectly with the error estimates of Table 3, providing a $RMS$ for the magnitude differences of 0.10 mag, and a zero point offset of 0.02 mag, with no appreciable dependence on the magnitude.

## 4. Spectroscopic Identifications and Individual Notes

The follow-up observations of the QSO candidates have been carried out at the 2.2m and 1.5m ESO telescopes at La Silla, equipped either with the ESO Faint Object Spectrograph and Camera (EFOSC2), or with the Boller and Chivens Spectrographs. The detectors were always CCDs. The resolution of the spectra ranged between 10 and 30 Å.

The reduction process used the standard MIDAS facilities (Banse et al. 1983) available at the Padova Department of Astronomy and at ESO Garching. The raw data were sky-subtracted and corrected for pixel-to-pixel sensitivity variations by division by a suitably normalized exposure of the spectrum of an incandescent source. Wavelength calibration was carried out by comparison with exposures of He-Ar, He, Ar and Ne lamps. Relative flux calibration was achieved by observations of standard stars listed by Oke (1974) and Stone (1977). A S/N ratio per resolution element larger than 10 was obtained for all the

**Table 4.** - continued

## Field SGP

| α (1950) | δ (1950) | $B_J$ | $U - B_J$ | $B_J - OR$ | Id. | z | Ref. |
|---|---|---|---|---|---|---|---|
| 0 40 32.95 | -30 24 09.6 | 17.93 | -0.89 | 0.09 | Q | 0.609 | 2 |
| 0 40 41.14 | -29 17 21.7 | 17.75 | -0.89 | 0.17 | Q | 2.084 | |
| 0 40 45.54 | -29 47 56.4 | 16.61 | -0.40 | 0.51 | s | 0.000 | |
| 0 40 46.05 | -29 19 39.4 | 18.37 | -0.77 | 0.06 | Q | 0.624 | 2 |
| 0 40 50.66 | -30 18 00.8 | 18.34 | -1.04 | 0.19 | Q | 0.496 | |
| 0 40 54.82 | -26 55 30.0 | 17.34 | -0.52 | 0.06 | Q | 1.002 | |
| 0 41 14.89 | -26 38 33.1 | 18.24 | -0.54 | 0.37 | Q | 3.053 | 2 |
| 0 41 23.02 | -27 21 21.5 | 18.35 | -0.45 | 0.54 | s | 0.000 | |
| 0 41 23.93 | -28 44 05.9 | 18.18 | -0.27 | 0.15 | Q | 0.839 | 2 |
| 0 41 28.39 | -29 04 15.9 | 17.80 | -0.61 | -0.03 | Q | 0.674 | 2 |
| 0 41 30.21 | -27 28 32.9 | 18.07 | -0.36 | 0.42 | s | 0.000 | |
| 0 41 30.83 | -26 07 38.3 | 17.30 | -0.78 | 0.36 | Q | 2.501 | 2 |
| 0 41 38.07 | -26 58 27.5 | 18.54 | -0.47 | 0.34 | Q | 2.457 | 2 |
| 0 41 38.59 | -26 25 52.0 | 18.68 | -0.61 | 0.30 | g | ... | |
| 0 41 40.60 | -28 59 35.5 | 17.97 | -0.80 | 0.12 | Q | 2.134 | 2 |
| 0 41 48.37 | -28 01 20.4 | 18.30 | -0.39 | 0.41 | g | ... | |
| 0 42 07.20 | -29 08 14.7 | 18.29 | -1.03 | 0.35 | Q | 1.250 | 2 |
| 0 42 16.86 | -25 50 32.4 | 18.27 | -0.38 | 0.15 | Q | 0.454 | 2 |
| 0 42 26.56 | -27 50 18.8 | 18.13 | -0.74 | -0.10 | Q | 0.741 | 2 |
| 0 42 32.88 | -26 22 04.2 | 18.57 | -0.58 | -0.38 | g | 0.108 | |
| 0 42 36.16 | -28 53 54.2 | 18.36 | -1.07 | -0.12 | s | 0.000 | |
| 0 42 40.13 | -27 13 41.8 | 18.15 | -0.39 | 0.43 | s | 0.000 | |
| 0 42 41.26 | -29 04 58.4 | 18.01 | -0.36 | 0.45 | s | 0.000 | |
| 0 42 41.68 | -29 30 57.9 | 17.92 | -0.48 | 0.40 | Q | 2.388 | 2 |
| 0 42 48.78 | -27 44 45.0 | 18.06 | -0.03 | -0.04 | g | ... | |
| 0 43 06.89 | -28 58 21.6 | 16.35 | -0.19 | -0.64 | s | 0.000 | |
| 0 43 22.32 | -27 43 31.8 | 18.46 | -0.70 | 0.35 | Q | 1.049 | 2 |
| 0 43 49.69 | -30 06 35.9 | 18.65 | -1.00 | 0.37 | Q | 1.124 | 2 |
| 0 43 55.71 | -26 16 37.6 | 18.27 | -1.11 | -0.28 | s | 0.000 | |
| 0 44 10.62 | -26 11 25.0 | 17.52 | -0.77 | 0.52 | Q | 0.129 | 2 |
| 0 44 22.49 | -29 51 22.4 | 18.12 | -0.44 | 0.48 | Q | 0.207 | 2 |
| 0 44 36.90 | -26 53 25.9 | 18.32 | -0.96 | 0.07 | s: | 0.000 | |
| 0 44 54.57 | -27 36 36.2 | 18.08 | -0.40 | 0.02 | s | 0.000 | |
| 0 45 04.49 | -30 02 53.7 | 18.27 | -1.05 | 0.25 | Q | 2.013 | |
| 0 45 14.09 | -28 18 50.7 | 17.79 | -0.52 | 0.07 | s | 0.000 | |
| 0 45 27.80 | -27 42 34.9 | 18.18 | -0.41 | 0.45 | Q | 0.232 | |
| 0 45 41.38 | -28 06 22.2 | 18.54 | -1.03 | 0.34 | Q | 1.138 | 2 |
| 0 45 45.22 | -26 06 23.5 | 18.08 | -0.73 | 0.27 | Q | 1.242 | 2 |
| 0 45 50.36 | -29 10 22.0 | 17.49 | -0.47 | 0.79 | s | 0.000 | |
| 0 45 59.48 | -25 45 18.0 | 18.08 | -1.29 | -0.51 | s | 0.000 | |
| 0 46 01.56 | -28 12 39.5 | 18.32 | -1.05 | 0.37 | Q | 1.687 | 2 |
| 0 46 16.70 | -27 20 17.9 | 16.71 | -0.17 | -0.11 | s | 0.000 | |
| 0 46 17.81 | -28 34 01.3 | 17.80 | -1.21 | 0.43 | Q | 0.632 | 2 |
| 0 46 33.19 | -29 59 25.0 | 18.14 | -0.59 | 0.06 | s | 0.000 | |
| 0 46 50.51 | -29 14 40.3 | 18.12 | -1.06 | 0.60 | Q | 0.781 | 2 |
| 0 47 03.86 | -26 36 34.9 | 17.72 | -0.47 | 0.54 | s | 0.000 | |
| 0 47 07.35 | -26 47 19.1 | 18.31 | -0.46 | 0.54 | s | 0.000 | |
| 0 47 18.05 | -26 30 13.9 | 17.49 | -0.27 | 0.35 | s | 0.000 | |
| 0 47 49.53 | -27 59 35.7 | 18.44 | -1.15 | 0.40 | Q | 2.143 | 2 |
| 0 47 54.17 | -26 47 53.3 | 18.62 | -0.79 | 0.18 | Q | 0.496 | 2 |
| 0 47 55.80 | -27 31 56.1 | 17.89 | -0.82 | -0.12 | s | 0.000 | |
| 0 48 05.01 | -26 14 03.4 | 18.41 | -0.14 | 0.12 | s | 0.000 | |

## Field SGP - continued

| α (1950) | δ (1950) | $B_J$ | $U - B_J$ | $B_J - OR$ | Id. | z | Ref. |
|---|---|---|---|---|---|---|---|
| 0 48 12.14 | -29 39 06.2 | 18.57 | -0.95 | -0.34 | s | 0.000 | |
| 0 48 13.96 | -25 57 40.4 | 18.69 | -0.70 | 0.32 | Q | 0.780 | 2 |
| 0 48 24.07 | -29 01 46.0 | 18.28 | -0.68 | 0.31 | Q | 0.783 | 2 |
| 0 48 25.80 | -28 42 25.8 | 17.87 | -0.66 | 0.10 | s | 0.000 | |
| 0 48 26.31 | -29 18 15.8 | 18.30 | -0.50 | 0.14 | Q | 0.428 | 2 |
| 0 48 31.72 | -30 05 46.0 | 15.85 | -0.39 | 0.88 | s | 0.000 | |
| 0 48 38.80 | -29 21 00.2 | 17.22 | -1.02 | -0.25 | s | 0.000 | |
| 0 48 40.38 | -29 51 47.5 | 16.77 | -0.38 | 0.56 | s | 0.000 | |
| 0 48 46.92 | -28 04 18.8 | 17.80 | -0.74 | 0.50 | Q | 0.846 | |
| 0 49 01.09 | -28 20 52.8 | 18.42 | -0.86 | 0.37 | Q | 2.249 | 2 |
| 0 49 08.57 | -26 58 15.4 | 16.98 | -1.20 | -0.28 | s | 0.000 | |
| 0 49 13.15 | -27 53 10.6 | 18.41 | -0.72 | 0.61 | Q | 0.410 | |
| 0 49 32.11 | -29 28 27.0 | 15.65 | -0.98 | -0.35 | s | 0.000 | |
| 0 49 42.21 | -28 40 27.5 | 18.68 | -0.48 | 0.12 | Q | 0.660 | 2 |
| 0 49 43.71 | -30 24 15.4 | 17.57 | -0.79 | 0.22 | Q | 0.471 | |
| 0 49 59.09 | -29 44 58.2 | 18.09 | -0.34 | 0.14 | HII | 0.114 | |
| 0 50 00.25 | -30 34 53.4 | 17.26 | -0.39 | 0.62 | s | 0.000 | |
| 0 50 05.78 | -29 05 36.7 | 18.39 | -0.45 | 0.24 | Q | 1.605 | 2 |
| 0 50 15.66 | -26 22 44.5 | 18.43 | -0.48 | 0.70 | HII | 0.176 | |
| 0 50 26.82 | -28 42 12.2 | 18.51 | -0.94 | 0.29 | Q | 1.650 | |
| 0 50 32.98 | -26 41 28.0 | 18.42 | -1.32 | 0.88 | Q | 1.248 | 2 |
| 0 50 36.87 | -29 29 13.4 | 18.57 | -0.54 | 0.13 | Q | 0.830 | 2 |
| 0 50 47.91 | -28 50 05.5 | 18.38 | -0.19 | 0.12 | g | ... | |
| 0 50 49.07 | -28 06 29.2 | 18.69 | -0.87 | 0.40 | Q | 1.805 | 2 |
| 0 50 50.87 | -27 42 05.8 | 17.70 | -0.65 | 0.04 | Q | 0.481 | 2 |
| 0 51 21.62 | -28 39 58.4 | 18.33 | -0.88 | 0.47 | Q | 1.574 | 2 |
| 0 51 42.62 | -28 01 13.1 | 18.51 | -0.47 | 0.65 | Q | 1.504 | |
| 0 51 48.39 | -26 30 46.0 | 18.09 | -0.05 | -0.02 | g | ... | |
| 0 51 48.57 | -30 12 03.1 | 18.69 | -1.13 | 0.67 | Q | 1.141 | 2 |
| 0 51 53.36 | -26 05 14.5 | 18.04 | -0.55 | 0.08 | Q | 0.624 | 2 |
| 0 52 00.72 | -30 20 54.4 | 17.55 | -0.80 | 0.24 | Q | 0.993 | 2 |
| 0 52 18.76 | -27 07 59.8 | 17.51 | -0.81 | -0.02 | s | 0.000 | |
| 0 52 26.88 | -29 45 50.0 | 18.58 | -0.35 | 0.09 | Q | 0.760 | 2 |
| 0 52 32.00 | -27 06 38.0 | 18.45 | -0.83 | -0.18 | s | 0.000 | |
| 0 52 33.40 | -27 46 33.7 | 17.21 | -0.71 | 0.09 | s | 0.000 | |
| 0 52 34.79 | -26 12 53.4 | 17.31 | -0.39 | 0.51 | s | 0.000 | |
| 0 52 40.87 | -28 56 49.4 | 18.33 | -0.54 | 0.34 | Q | 0.602 | 2 |
| 0 52 41.10 | -25 50 36.4 | 18.51 | -0.72 | -0.07 | Q | 0.855 | 2 |
| 0 52 51.31 | -28 53 35.4 | 18.65 | -0.51 | 0.24 | Q | 0.634 | 2 |
| 0 53 11.28 | -26 39 00.7 | 18.44 | -0.80 | 0.46 | Q | 0.808 | 2 |
| 0 53 38.52 | -27 09 10.5 | 18.02 | -0.43 | 0.28 | Q | 1.039 | 2 |
| 0 53 47.14 | -28 13 24.5 | 18.44 | -0.36 | 0.06 | Q | 0.725 | 2 |
| 0 53 53.73 | -26 53 25.6 | 16.33 | -0.16 | -0.09 | s | 0.000 | |
| 0 54 04.13 | -27 49 18.8 | 18.18 | -0.51 | 0.44 | s | 0.000 | |
| 0 54 21.80 | -30 29 23.1 | 18.20 | -0.86 | -0.16 | s | 0.000 | |
| 0 54 38.29 | -30 17 44.2 | 18.18 | -0.10 | -0.05 | g | ... | |
| 0 54 46.00 | -28 42 06.5 | 18.34 | -0.51 | 0.57 | s | 0.000 | |
| 0 54 46.44 | -30 14 58.4 | 17.10 | -0.50 | 0.75 | s | 0.000 | |
| 0 54 58.47 | -29 58 04.7 | 17.59 | -0.85 | -0.24 | s | 0.000 | |
| 0 55 09.64 | -27 44 40.2 | 18.41 | -0.40 | 0.31 | Q | 2.174 | 2 |
| 0 55 14.34 | -25 59 04.6 | 18.04 | -0.67 | 0.23 | Q | 0.584 | 2 |
| 0 55 24.37 | -27 46 28.6 | 18.40 | -1.19 | 0.81 | g | ... | |

## Field SGP - continued

| α (1950) | δ (1950) | $B_J$ | $U - B_J$ | $B_J - OR$ | Id. | z | Ref. |
|---|---|---|---|---|---|---|---|
| 0 55 32.41 | -29 45 36.9 | 17.09 | -0.26 | 0.31 | s | 0.000 | |
| 0 55 39.83 | -28 41 21.8 | 18.38 | -0.55 | 0.70 | s | 0.000 | |
| 0 55 43.39 | -29 49 00.3 | 18.58 | -0.79 | 0.34 | Q | 0.668 | 2 |
| 0 55 59.10 | -28 52 57.7 | 18.54 | -0.73 | -0.23 | s | 0.000 | |
| 0 56 14.07 | -28 40 37.8 | 18.56 | -0.79 | 0.05 | s | 0.000 | |
| 0 56 15.43 | -29 46 10.5 | 17.51 | -0.69 | 1.04 | nq | ... | |
| 0 56 15.97 | -26 25 23.1 | 17.14 | -0.43 | 0.63 | s | 0.000 | |
| 0 56 20.86 | -29 15 35.3 | 18.56 | -1.01 | 0.53 | Q | 1.255 | 2 |
| 0 56 25.92 | -27 58 45.3 | 18.66 | -0.66 | 0.40 | s | 0.000 | |
| 0 56 26.51 | -29 45 59.5 | 17.05 | -0.45 | 0.86 | s | 0.000 | |
| 0 56 31.92 | -27 33 51.3 | 18.28 | -0.03 | -0.05 | s | 0.000 | |
| 0 56 41.38 | -28 43 14.6 | 17.69 | -0.84 | 0.37 | Q | 0.934 | 2 |
| 0 56 58.14 | -29 48 19.7 | 18.09 | -0.11 | 0.14 | Q | 0.351 | 2 |
| 0 57 03.44 | -26 47 11.6 | 16.36 | -0.41 | 0.77 | s | 0.000 | |
| 0 57 10.23 | -30 20 06.2 | 18.38 | -0.54 | 0.29 | Q | 0.394 | |
| 0 57 13.26 | -26 57 18.1 | 18.18 | -0.48 | 0.65 | s | 0.000 | |
| 0 57 21.21 | -29 55 09.3 | 18.31 | -0.44 | 0.40 | s | 0.000 | |
| 0 57 29.01 | -30 03 55.5 | 17.54 | -0.50 | 0.59 | s | 0.000 | |
| 0 57 29.11 | -26 10 48.6 | 16.89 | -0.60 | 0.13 | s | 0.000 | |
| 0 57 30.20 | -26 30 21.6 | 18.58 | -0.79 | 0.40 | Q | 1.042 | 2 |
| 0 57 40.73 | -29 08 15.8 | 18.32 | -0.87 | 0.37 | Q | 0.489 | 2 |
| 0 57 45.74 | -28 22 37.1 | 17.52 | -0.91 | -0.14 | s | 0.000 | |
| 0 57 50.75 | -27 35 44.8 | 18.28 | -0.27 | 0.22 | g | ... | |
| 0 57 51.24 | -27 12 59.6 | 18.35 | -0.49 | 0.65 | s | 0.000 | |
| 0 57 56.68 | -27 29 45.9 | 18.43 | -0.92 | 0.58 | Q | 1.203 | 2 |
| 0 57 57.49 | -26 26 05.9 | 18.52 | -0.79 | -0.01 | s | 0.000 | |
| 0 58 05.23 | -26 18 27.9 | 18.06 | -0.67 | 0.27 | Q | 0.792 | |
| 0 58 06.24 | -26 04 43.9 | 18.50 | -0.69 | 0.48 | Q | 2.472 | 2 |
| 0 58 14.03 | -25 54 35.7 | 16.73 | -0.17 | 0.17 | Q | 0.158 | 2 |
| 0 58 28.42 | -27 09 28.9 | 18.04 | -0.50 | 0.69 | s | 0.000 | |
| 0 58 33.43 | -26 58 54.2 | 18.45 | -0.49 | 0.60 | s | 0.000 | |
| 0 58 34.57 | -29 19 01.6 | 18.27 | -0.73 | 0.40 | Q | 1.184 | 2 |
| 0 58 35.72 | -29 07 24.0 | 18.00 | -0.77 | 0.32 | Q | 0.866 | 2 |
| 0 58 40.25 | -25 59 10.4 | 18.34 | -0.52 | 0.67 | s | 0.000 | |
| 0 58 41.44 | -28 31 00.8 | 18.62 | -0.69 | 0.37 | Q | 1.001 | 2 |
| 0 58 41.91 | -26 20 49.4 | 18.04 | -0.83 | 0.38 | g | ... | |
| 0 58 43.87 | -27 02 15.4 | 18.02 | -0.75 | 0.13 | s | 0.000 | |
| 0 59 08.22 | -27 29 03.1 | 17.78 | -0.73 | 0.77 | Q | 0.188 | 2 |
| 0 59 10.71 | -28 53 38.7 | 18.14 | -0.29 | 0.11 | Q | 0.620 | 2 |
| 0 59 26.42 | -28 10 45.6 | 16.97 | -0.46 | 0.64 | s | 0.000 | |
| 0 59 27.79 | -26 25 06.4 | 18.51 | -0.91 | 0.13 | Q | 2.109 | 2 |
| 0 59 37.26 | -30 34 35.6 | 16.92 | -0.51 | 0.28 | Q | 1.033 | 2 |
| 0 59 39.23 | -30 13 26.7 | 17.33 | -0.65 | 0.02 | s | 0.000 | |
| 0 59 50.40 | -29 46 20.1 | 17.73 | -0.85 | 0.28 | Q | 1.076 | 2 |
| 0 59 57.15 | -26 57 53.4 | 18.62 | -0.71 | 0.27 | Q | 2.266 | 2 |
| 1 00 16.07 | -26 31 02.0 | 18.70 | -1.09 | -0.06 | s | 0.000 | |
| 1 00 22.52 | -28 03 19.5 | 18.37 | -0.53 | 0.61 | s | 0.000 | |
| 1 00 27.64 | -28 09 09.7 | 18.04 | -0.49 | 0.86 | Q | 1.768 | 2 |
| 1 00 31.70 | -27 02 42.5 | 17.78 | -1.06 | 0.32 | Q | 1.597 | 2 |
| 1 00 32.78 | -28 51 15.7 | 17.50 | -0.60 | 0.78 | s | 0.000 | |
| 1 00 35.75 | -27 14 33.9 | 17.91 | -0.58 | -0.08 | s | 0.000 | |
| 1 00 44.02 | -28 07 51.5 | 17.31 | -0.31 | 0.33 | s | 0.000 | |

**Field SGP - end**

| $\alpha$ (1950) | $\delta$ (1950) | $B_J$ | $U - B_J$ | $B_J - OR$ | Id. | $z$ | Ref. |
|---|---|---|---|---|---|---|---|
| 1 00 47.08 | -30 30 54.0 | 18.66 | -0.84 | 0.40 | Q | 0.940 | |
| 1 00 52.13 | -26 49 55.9 | 17.34 | -0.19 | 0.13 | HII | 0.156 | |
| 1 00 59.33 | -27 02 15.4 | 16.96 | -0.91 | 0.27 | s | 0.000 | |
| 1 01 00.35 | -29 15 36.1 | 16.61 | -0.24 | 0.46 | s | 0.000 | |
| 1 01 01.90 | -27 08 49.7 | 18.50 | -0.76 | 0.00 | Q | 0.558 | 2 |
| 1 01 04.53 | -28 21 56.2 | 18.63 | -0.85 | -0.21 | s | 0.000 | |
| 1 01 08.34 | -25 48 31.0 | 18.58 | -1.00 | 0.17 | Q | 1.973 | 2 |
| 1 01 14.33 | -29 10 17.2 | 17.05 | -0.39 | 0.79 | s | 0.000 | |
| 1 01 19.27 | -27 07 58.6 | 15.00 | -1.23 | -0.47 | s | 0.000 | |
| 1 01 23.90 | -26 11 05.0 | 17.84 | -0.36 | 0.54 | Q | 0.277 | 2 |
| 1 01 27.07 | -29 46 09.8 | 17.91 | -0.54 | 0.62 | s | 0.000 | |
| 1 01 56.45 | -29 38 18.2 | 18.31 | -0.50 | 0.44 | g | ... | |
| 1 02 16.71 | -27 13 11.4 | 17.33 | -0.68 | 0.38 | Q | 0.780 | 2 |
| 1 02 35.42 | -30 23 25.0 | 17.64 | -0.45 | 0.51 | s | 0.000 | |
| 1 02 38.79 | -28 42 30.3 | 18.65 | -0.92 | 0.53 | Q | 1.375 | |
| 1 02 43.16 | -30 12 01.0 | 17.78 | -0.60 | 0.15 | Q | 0.838 | 2 |
| 1 02 59.53 | -30 14 53.1 | 18.24 | -0.62 | 0.09 | Q | 0.533 | |

QSO candidates and the spectral range was always extended enough to allow an unquestionable identification also in the (few) cases in which only one line was visible in the spectrum (assumed to be 2798Å MgII).

In Table 4 the complete list of the candidates and their spectroscopic identification is given field by field. The identification classes are: $Q$ = QSO or Seyfert 1 galaxy; $HII$ = starburst galaxy; $s$ = star; $g$ = galaxy; $ext$ = very extended object. Uncertain identifications and redshifts are indicated with a colon (:). In some cases a negative identification is reported with the prefix $n$: for example $nq$ means that the object is "not a QSO". The area of a field may overlap other fields, in this case each object in common with another field is marked with the overlapping field number in the column $Ov.$ (overlap).

## 4.1. F287

The search in this field has been carried out over an area limited by a rectangle in $\alpha$ and $\delta$ rather than in $RAC$ and $\delta$ as for all the other fields. For this reason in Table 1 we have reported in columns 4 and 5 the $\alpha_{min}$ and $\alpha_{max}$ limits in parentheses. The list of the candidates has been obtained from the database described by Hawkins and Véron (1993). Due to the high surface density of objects (the galactic latitude of this field is $-47°$), a rela-tively large incompleteness (about 10%) is present because of the blending of the QSO images with nearby objects. Two more QSOs, in fact, are reported in Table 5 that have been missed because their images were merged with those of a faint star or galaxy.

## 4.2. F295

Two portions of the field have been excluded from the search. The first has rectangular shape and is delimited by the points ($RAC = 0.2764$, $\delta = -37°46'46''$; north-west) and ($RAC = 0.6069$, $\delta = -38°05'21''$; south-east). The second is in the lower right corner, with a triangular shape, delimited by the vertices ($RAC = 2.17415$, $\delta = -42°38'10''$; south-west), ($RAC = 2.61153$, $\delta = -42°12'58''$; north-east) and ($RAC = 2.61153$, $\delta = -42°38'10''$; field corner). One QSO has been missed (Q0044-3736, z=1.233) because its image is very close to another point-like object and was not measured on the J plates. The candidate has been observed by us thanks to a previous selection carried out on the basis of different photographic material. The QSO lies in the region overlapping with F351 and has been missed for the same reason in that field too. The QSOs 0042-3249, 0046-4143, 0046-3828, 0049-3847, 0102-3732 have been ob-

reported (as probable) in the Véron catalog.

### 4.3. F351

A portion of the field has been excluded from the search. Its shape is rectangular and is delimited between $\delta = -34°12'40''$ and $\delta = -33°44'27''$ with $RAC$ greater than 2.007. One QSO has been missed (Q0044-3736, z=1.233) because its image is very close to another point-like object and was not measured on the U plates. The QSO lies in the region overlapping with F295 and has been missed for the same reason in that field too (see previous paragraph). The QSO Q0055-3355 (z=0.83), reported by Kunth & Sargent (1986) with approximate coordinates, has been re-discovered with coordinates $\alpha = 00^h55^m49^s.44, \delta = -33°54'02''.7$ (z=0.832)

### 4.4. F534

Two portions of the field have been excluded from the search. They correspond to the sensitometric spots and are of rectangular shape. The first zone lies between $\delta = -25°50'32''$ and $\delta = -24°48'50''$ with $RAC$ lower than -2.303, the second lies between $\delta = -25°08'54''$ and $\delta = -24°07'13''$ with $RAC$ larger than 2.380.

### 4.5. SA94

A portion of the field has been excluded from the search. It corresponds to the south-east corner, with a triangular shape, delimited by the vertices ($RAC = 2.390$, $\delta = -2°46'45''$), ($RAC = 2.713$, $\delta = -2°2'53''$) and ($RAC = 2.713$, $\delta = -2°46'45''$, the field corner).

One QSO has been missed: Q0257+0229 is a low-redshift, low-luminosity ($M_B = -23.08$) object with rather extended images on our plates. Its measured $U - B$ colour ($U - B = 0.19$) was too red to meet the selection criterion.

### 4.6. SGP

A portion of the field has been excluded from the search because of the presence of NGC288. Its shape is circular with center at $\alpha = 00^h50^m17^s.4, \delta = -26°51'25''$ and a radius of 0.15 deg. Three QSOs have been missed (see Table 5). Q0049-2939 has its image merged with that of a nearby star. Q0059-2545 lies just outside the selection criterion ($U - J = -0.30$, $J - R = 0.41$) either because of the photometric errors or due to intrinsic redder-than-average colours. Q0059-2735 is a BAL (Morris et al. 1991) and has intrinsically red colours ($U - J = 0.04$, $J - R = 0.81$).

The assessment of the completeness of UVx surveys for low-redshift QSOs has been made by La Franca et al. (1992). The incompleteness in this magnitude interval with the present accuracy is expected to be about 5%, due to photometric errors and intrinsic spread in the QSO colors. Additional sources of incompleteness (see e.g. the individual notes on the missed QSOs) are due to overlapping images, spectral peculiarities (e.g. broad absorption lines), inaccurate photometry of the extended objects.

QSOs known from the literature that should have been selected in the present survey but have been missed are reported in Table 5. Out of 154 previously known QSOs with z < 2.2, 8 have been missed, confirming the above estimate of the incompleteness.

**Table 5.** Missed QSOs

| Field | $\alpha$ (1950) | $\delta$ (1950) | $B_J$ | z | Ref. |
|-------|-----------------|-----------------|-------|-------|------|
| 287   | 21 35 14.40     | -45 30 49.7     | 17.7  | 0.308 | 1    |
| 287   | 21 35 14.42     | -42 43 47.4     | 17.2  | 0.250 | 1    |
| 295   | 00 44 40.65     | -37 36 12.1     | 18.2  | 1.233 |      |
| 351   | 00 44 40.65     | -37 36 12.1     | 18.2  | 1.233 |      |
| SA94  | 02 57 53.90     | +02 29 01.0     | 16.1  | 0.115 | 2    |
| SGP   | 00 49 12.70     | -29 39 31.0     | 17.6  | 0.308 | 2    |
| SGP   | 00 59 21.44     | -25 45 56.9     | 18.2  | 1.955 | 2    |
| SGP   | 00 59 52.54     | -27 35 56.6     | 18.0  | 1.595 | 2    |

Only two other surveys covering with statistical significance the range $17 < B < 18.5$ exist at present in the literature: the Edinburgh QSO survey (Goldschmidt et al. 1991) and the LBQS (Hewett et al. 1993).

It is therefore of interest to compare the properties of our selection (very similar to the Edinburgh survey) with the LBQS (an objective-prism survey). Three of the presently discussed fields (F287, SGP, SA94) partially or totally overlap with corresponding LBQS fields. The photometric calibrations (all carried out in the $B_J$ band) are in good agreement: in F287 no zero point difference has been found while the $RMS$ of the magnitude differences is 0.14 mag (a pessimistic estimate considering that it has been obtained comparing the QSOs in common between the two samples, expected to be at least in part variable); in the SGP a zero point offset (LBQS – HBQS magnitudes) of 0.07 mag has been measured with an $RMS$ of 0.11 mag; in the SA94 a zero point offset of −0.13 mag has been measured with an $RMS$ of 0.15 mag.

On the basis of these relationships we have searched the LBQS database for all the HBQS QSOs (with z > 0.3) brighter than $B_J = 18.25$, 18.45 and 18.30 in the F287, SGP and SA94, respectively (corresponding to a $2\sigma$ limit

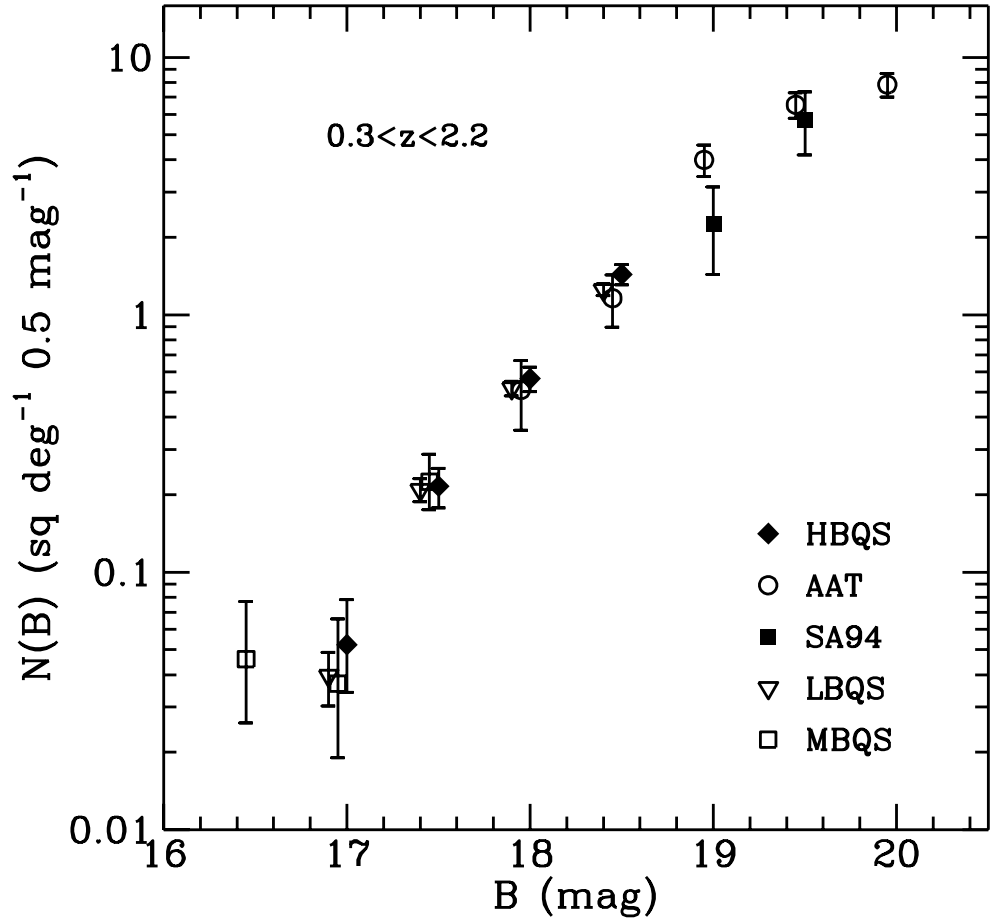

**Fig. 2.** QSO differential counts as a function of the $B$ magnitude. The results of the present work are compared with the MBQS (Mitchell et al. 1984), LBQS (Hewett et al. 1993), SA94 (La Franca et al. 1992) and AAT (Boyle et al. 1990) samples. The abscissae of the points referring to the AAT and MBQS samples have been translated of $-0.05$ mag, for the sake of graphical clarity. Note that the LBQS counts have been computed in bins that are shifted of $-0.1$ mag with respect to Table 6

**Table 6.** QSOs counts and surface densities for $M_B < -23$ and $0.3 < z \leq 2.2$

| Magnitude range $B$ | HBQS N. | HBQS Dens. $deg^{-2}\ 0.5\ mag^{-1}$ | LBQS N. | LBQS Dens. $deg^{-2}\ 0.5\ mag^{-1}$ | AAT N. | AAT Dens. $deg^{-2}\ 0.5\ mag^{-1}$ |
|---|---|---|---|---|---|---|
| 16.75 - 17.25 | 8 | $0.05^{0.08}_{0.03}$ | 29 | $0.06^{0.08}_{0.05}$ | — | — |
| 17.25 - 17.75 | 33 | $0.22^{0.25}_{0.18}$ | 122 | $0.27^{0.29}_{0.24}$ | — | — |
| 17.75 - 18.25 | 85 | $0.56^{0.62}_{0.50}$ | 255 | $0.58^{0.61}_{0.54}$ | 11 | $0.51^{0.66}_{0.36}$ |
| 18.25 - 18.75 | 128 | $1.44^{1.56}_{1.31}$ | 328 | $1.38^{1.46}_{1.31}$ | 19 | $1.16^{1.43}_{0.89}$ |

ness limits for these three fields). It resulted that out of
82 QSOs (16,54,12 in the three fields) 16 (5,10,1) have
not been found in the LBQS. However, of the 16 miss-
ing objects, 12 (Hewett, private communication) turn out
to be absent from the LBQS because they are in step-
wedges or regions close to bright stars, or they have un-
processable objective-prism spectra due to overlaps. Only
3, although with useful objective-prism spectra, have not
been included in the published lists. This corresponds to
an incompleteness of the LBQS that is at most 4%.

## 6. The QSO Counts

The number magnitude relation from the HBQS sam-
ple in the redshift interval $0.3 < z \leq 2.2$ and absolute
magnitudes $M_B \leq -23$ is shown in Fig. 2 and Table 6.
All the missed QSOs of Table 5 in the redshift interval
$0.3 < z \leq 2.2$ have been included. No other correction for
incompleteness has been taken into account.

In Fig. 2 and Table 6 our results are compared
with those of the MBQS (Mitchell et al. 1984), LBQS
(Hewett et al. 1993), SA94 (La Franca et al. 1992), and
AAT (Boyle et al. 1990) samples. The absolute mag-
nitudes have been computed applying the K corrections
on the basis of the composite spectrum by Cris-
tiani & Vio (1990). We have adopted the value $H_0 =
50\ Km\ s^{-1} Mpc^{-1}$ for the Hubble parameter and $q_0 = 0.5$.
For the AAT sample the counts published in Table 9 of
Boyle et al. (1990) have been used. The apparent magni-
tudes of the objects of the HBQS and LBQS samples have
been transformed to the Cousins B photometric passband
according to

$$B = B' + 0.11(B - V) \qquad (1)$$

$$B = B_J + 0.23(B - V) \qquad (2)$$

where $B'$ and $B_J$ are the "natural" photographic mag-
nitudes with the IIaO + GG385, and IIIaJ + GG395
emulsions and filters respectively (Blair & Gilmore 1982).
An average $B - V = 0.3$ color has been assumed for all the
QSOs, and dereddening for galactic extinction according
to Burstein & Heiles (1982) has been applied.

In Fig. 3 the histogram with the redshift distribution
is shown. The shape is typical of a UVx survey with a
relatively bright flux limit, for which the fraction of lower-
redshift QSOs is larger with respect to fainter samples.

As shown in Fig. 2, the counts in the magnitude in-
terval $17 < B < 19$ are now firmly defined. Indeed, in
this magnitude range the HBQS and LBQS surveys show
a good agreement. This is not in contrast with the dis-
crepancies reported in the previous section, since for the
LBQS counts the incompleteness has been taken into ac-
count by using the effective area of each subsample. A
certain degree of incompleteness may also be masked by

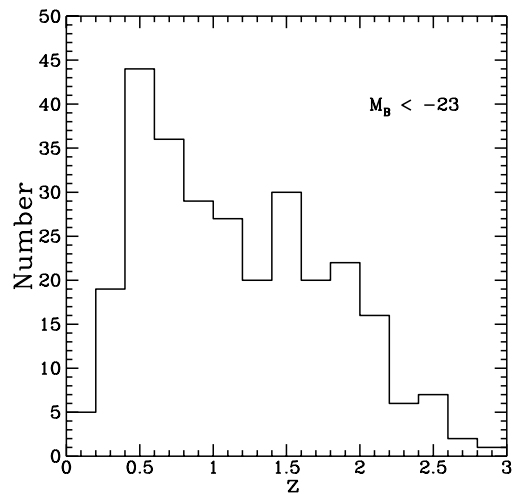

**Fig. 3.** Number-redshift histogram for the 285 QSOs
($M_B < -23$) of the HBQS sample

slight shifts in the zero points of the photometric calibra-
tions. Besides, the color transformations 1), 2) may not
allow accurate comparisons for QSOs.

With the addition of the bright part of the HBQS we
expect to further decrease the uncertainty in the interval
$17.0 < B < 17.5$, and to achieve a very reliable measure-
ment of the $0.3 < z < 2.2$ QSO counts in the magnitude
interval $15 < B < 17$.

*Acknowledgements.* It is a pleasure to thank M. R. S. Hawkins
for providing us with data for the field 287. This work would
have not been possible without the enthusiastic support of
the COSMOS and UKST staff both in Edinburgh and Coon-
abarabran. This work has been partially supported by the
ASI contracts 92-RS-102 and 94-RS-107 and from the EC pro-
gramme Human Capital and Mobility. One of us, RV, thanks
an ESA research fellowship.